\documentclass[12pt, preprint]{aastex}
\usepackage{graphicx, subfigure, lscape}
\usepackage{natbib}
\slugcomment{Accepted for publication in The Astronomical Journal}

\shorttitle{The Solar Neighborhood {\it XXVIII}}
\shortauthors{Dieterich et al.}
\begin{document}

\title{The Solar Neighborhood {\it XXVIII}: The  Multiplicity Fraction of Nearby Stars from 5 to 
70 AU and the Brown Dwarf Desert Around M Dwarfs }

\author{Sergio B. Dieterich, Todd J. Henry}
\affil{Georgia State University, Atlanta, GA 30302-4106}
\email{dieterich@chara.gsu.edu}

\author{David A. Golimowski}
\affil{Space Telescope Science Institute, Baltimore, MD 21218}

\author{John E. Krist}
\affil{Jet Propulsion Laboratory, Pasadena, CA 91109}

\author{Angelle M. Tanner}
\affil{Mississippi State University, Starkville, MS 39762}

\begin{abstract}
We report on our analysis of {\it HST/NICMOS} snapshot high resolution
images of 255 stars in 201 systems within $\sim$10 parsecs of the Sun.
Photometry was obtained
through filters {\it F110W, F180M, F207M,} and {\it F222M} using {\it
NICMOS} Camera 2.  These filters were selected to permit clear
identification of cool brown dwarfs through methane contrast
imaging. With a plate scale of 76 {\it mas/pixel}, {\it NICMOS} can
easily resolve binaries with sub-arcsecond separations in the
19\farcs5$\times$19\farcs5 field of view. We previously reported five
companions to nearby M and L dwarfs from this search. No new
companions were discovered during the second phase of data analysis
presented here, confirming that stellar/substellar binaries are
rare. We establish  magnitude and separation limits for which
companions can be ruled out for each star in the sample, and then
perform a comprehensive sensitivity and completeness analysis for the
subsample of 138 M dwarfs in 126 systems.  We calculate a multiplicity
fraction of $0.0_{-0.0}^{+3.5}$\% for L companions to M dwarfs in
the separation range of 5 to 70 AU, and $2.3_{-0.7}^{+5.0}$\% for
L and T companions to M dwarfs in the separation range of 10 to 70
AU. We also discuss trends in the color-magnitude diagrams using
various color combinations and present astrometry for 19 multiple
systems in our sample. Considering these results and results from
several other studies, we argue that the so-called ``brown dwarf
desert'' extends to binary systems with low mass primaries and is
largely independent of primary mass, mass ratio, and
separations. While focusing on companion properties, we discuss how
the qualitative agreement between observed companion mass functions
and initial mass functions suggests that the paucity of brown dwarfs
in either population may be due to a common cause and not due to
binary formation mechanisms.

\end{abstract}


\clearpage

\section{Introduction}
\label{sec:intro}

The mass function, multiplicity fraction, and the mass-luminosity
relation are three of the most important characteristics of a stellar
or substellar population. However, all three remain poorly constrained
for Very Low Mass (VLM) stars. Although the lowest mass stars, the M
dwarfs, dominate the Galaxy in numbers and comprise the majority of
our stellar neighbors \citep{Henryetal2006}, not a single M dwarf is
visible to the naked eye. Over the last two decades, advances in
observational astronomy have made a thorough study of these faint
stars possible. Empirical mass-luminosity relations
\citep{HenryandMcCarthy1993,Henryetal1999,Delfosseetal2000} have
achieved a high degree of reliability for early to mid M dwarfs, with
progress continuing for later M dwarfs at the end of the main
sequence. Large sky surveys such as the 2 Micron All Sky Survey ({\it
2MASS}) in the near infrared \citep{Skrutskieetal2006} and the Sloan
Digital Sky Survey ({\it SDSS}) in the optical \citep{Yorketal2000}
have provided a wealth of new data for population studies, but lack
the angular resolution necessary to investigate the Multiplicity
Fraction (MF) and Companion Mass Function (CMF) at separations
corresponding to the short periods necessary for determining dynamical
masses.

The discovery of GJ 229B, the first unequivocal brown dwarf
\citep{Nakajimaetal1995}, followed by hundreds of others\footnote{A
current list of known L and T dwarfs is maintained at
www.dwarfarchives.org.}, raised fundamental questions about our
understanding of low mass star formation. Are VLM stars and brown
dwarfs products of a single mechanism of (sub)stellar formation applicable to
a wide range of masses?  Or, do brown dwarfs constitute a
fundamentally different population? Does the trend in the stellar mass
function producing more stars at lower masses continue into the brown
dwarf regime, therefore making them more numerous than their stellar
cousins? What do multiplicity properties, such as the overall
multiplicity fraction and the separation distribution, tell us about
the environments in which VLM stars and brown dwarfs were born? These
are some of the fundamental questions that have only recently been
addressed through a combination of sky surveys
\citep[e.g.,][]{Bochanskietal2010}, wide separation common proper
motion searches \citep[e.g.,][]{Allenetal2007,AllenandReid2008}, high
resolution multiplicity surveys
\citep[e.g.,][]{ReidandGizis1997,Closeetal2003,Gizisetal2003,Lowranceetal2005,Reidetal2008,MetchevandHillenbrand2009}
and the establishment of trigonometric parallaxes for a large sample
of objects \citep[e.g.,][]{Dahnetal2002,Henryetal2006}

Thorough characterization of any stellar population requires the study
of a volume limited sample. In an effort to better understand these
properties, we have conducted an {\it HST/NICMOS} snapshot program
imaging 255 objects in 201 star systems with trigonometric
parallaxes placing them within $\sim$10 pc of the Sun (Table 1). We
used the technique of methane imaging \citep{Rosenthaletal1996,Tinneyetal2005} 
to clearly distinguish cool brown dwarf
companions. In 2004 we reported the detection of four M dwarf
companions and one binary L dwarf in a triple system \citep[hereafter
G04]{Golimowskietal2004b}. With small infrared contrasts ranging from
$\sim$0 magnitudes (GJ 1001BC) to 4.5 magnitudes (GJ 84AB), the
companions we reported in 2004 were relatively bright. We have since
carried out a deeper search of the data, establishing formal
sensitivity limits for the detection of companions in the field of
each primary target and extending the limiting magnitude differences
routinely to 11 at separations of 3\farcs0, 8 at 1\farcs0, 4 at
0\farcs4, and 2 at 0\farcs2 ($\S$5.3, Figure 6a).  Having completed
the deeper search of the data with no further detections, we now
report on the magnitude and separation limits to which we can rule out
companions for each object in our sample. We also discuss what the
lack of additional brown dwarf detections tells us about the
multiplicity fraction of systems with VLM secondaries in these mass
and separation regimes.

We describe the general characteristics of our sample and discuss how
the observed sample relates to our current knowledge of the solar
neighborhood in $\S$2. Instrumental aspects of the observations
relevant to obtaining our sensitivity limits are briefly reviewed in
$\S$3, and a detailed discussion of our Point Spread Function (PSF)
insertion method for testing the sensitivity of the search is given in
$\S$4. We discuss photometric trends in our color system and note
several benchmark objects in $\S$5.1. We report new astrometric data
for 19 known binary systems in $\S$5.2, and discuss the sensitivity of
the search in $\S$5.3. In $\S$5.4 we establish a sub-sample of 126 M
dwarf systems for which we calculate the multiplicity fraction,
including substellar companions, from 5 to 70 AU based on companion
detections (or lack thereof) and completeness arguments.  We discuss
what our results mean in the context of the ``brown dwarf desert'' in
$\S$6.1 and $\S$6.2. Finally, we end by comparing young cluster
multiplicity studies, estimates of the Galactic disk luminosity
function, and our results in $\S$6.3, and summarize our conclusions in
$\S$7.

\section{General Characteristics of the Sample}
Our target list was designed to provide a sample that is
representative of the solar neighborhood. Because this survey was an
{\it HST} snapshot program, our targets were pulled from a large pool
of selected targets in order to fill small gaps in {\it HST's}
observing schedule. The REsearch Consortium on Nearby Stars
{\it(RECONS)}\footnote{Information about {\it RECONS} and periodically
updated versions of the 10 pc census are available at
www.recons.org. } is engaged in an effort to obtain a census and
thorough characterization of the population of stars within 25 pc of
the Sun, with a particular concentration on stars closer than 10 pc
\citep{Henryetal2006}.  In order to be a member of the {\it RECONS} 10
pc sample, an object must have a trigonometric parallax greater than
100 mas, with an error smaller than 10 mas. We used the {\it RECONS}
10 pc sample as a starting list for our search and allowed the {\it
HST} snapshot scheduling process to effectively select a random
subsample from the 10 pc sample.  Table 1 summarizes several tallies
of the observed sample.  These observations comprise 69\% of the {\it
RECONS} 10 pc sample (epoch 2012.0), including main sequence stars,
white dwarfs, L and T dwarfs, but excluding extrasolar planets. We
note that because trigonometric parallaxes for nearby stars are
constantly being updated, 17 objects in 13 systems originally included
in our search are no longer members of the 10 pc sample. We still
include their data as individual stars in this paper, but exclude them
from statistical considerations in order to keep the sample volume
limited.

Figure 1 shows the spectral type distribution of the {\it NICMOS}
snapshot sample. Out of the 218 resolved objects within 10 pc we
observed, 138, or 63\%, are M dwarfs. This number is a very close
match to the M dwarf fraction in the {\it RECONS} 10 pc sample, which
is 248 out of 357 objects, or 69\% (epoch 2012.0).  The preponderance
of M dwarfs in our sample means that even though the sample is a
random representation of the solar neighborhood, it focuses on the
spectral type that is least scrutinized by RV companion searches and
open cluster imaging searches. By studying nearby M dwarfs, which
comprise a disk rather than cluster population, we are mapping the
brown dwarf desert in a largely unexplored region.

\section{Observations and Data Reduction}

G04 describe technical aspects of the observations in detail. We give
a brief summary here and highlight the aspects that are most relevant
in achieving the sensitivities we later quote for each individual
target.

We obtained direct images of each target using {\it NICMOS} Camera 2
{\it (NIC2)} through four near infrared filters during cycles 7
(1997$-$1998) and 11 (2002$-$2003).  {\it NIC2} has a plate scale of
0\farcs076 pixel$^{-1}$ and a field of view of
19\farcs5$\times$19\farcs5 (Viana et al. 2009, Thatte et
al. 2009)\footnote{ {\it HST/NICMOS} documentation, including the {\it
NICMOS Instrument Handbook} and the {\it NICMOS Data Handbook}, is
currently available from the Space Telescope Science Institute at
www.stsci.edu/hst/nicmos}.  Targets were imaged through the {\it
F110W}, {\it F180M}, {\it F207M}, and {\it F222M} filters, centered at
1.10\micron, 1.80\micron, 2.07\micron, and 2.22\micron,
respectively. Because {\it HST} observations are not subject to atmospheric
absorption, the {\it NICMOS} filters are not defined to sample
atmospheric transmission windows in the way that ground based near
infrared filters are. The resulting filter set is non-standard when
compared to ground based systems, but allows the user to construct a
color scheme that is more suitable for the underlying physics being
investigated.  Figure 2 shows the transmission curves for the selected
{\it NICMOS} filters, with the {\it 2MASS J, H,} and {\it K$_S$}
filters also plotted for comparison.  The four filters in this survey
were selected to detect the strong CH$_4$ absorption bands observed in
T dwarf spectra at 1.7$\micron$ and 2.2$\micron$, in effect imaging in
and out of these absorption bands. Depending on the filter choice,
there is a drastic color shift of up to three magnitudes for T
dwarfs. A late T dwarf appears blue in {\it F110W$-$F180M} (0.0 to
$-$1.0, Figure 3a) whereas it is red in {\it F180M$-$F207M}
(1.0 to 2.0, Figure 3c). Because no background source is likely to
have such a strong color shift, T dwarfs are readily identified in
this color scheme. This technique is commonly known as methane
imaging and has been used to successfully identify brown
dwarfs in photometric observations \citep{Rosenthaletal1996,Tinneyetal2005}.

By centering the targets on the detector, we searched for companions
within a radius of 9\farcs5, except for a small ($\sim$1$\arcsec$ in
diameter) artifact due to the coronagraphic hole on the upper left
quadrant of the detector\footnote{{\it HST's} roll orientation during
a given exposure is constrained by the need to keep the solar arrays
facing the Sun. Consequently, the position angle of the coronagraphic
hole with respect to celestial north, as well as the image's overall
orientation, varies widely among the images of our targets.}.  A few
targets had large coordinate uncertainties, in most cases due to
poorly constrained high proper motions. These targets were not
properly centered in the field of view, and are specified in the notes
to Table 2. Although some of our primary targets are very bright
(e.g. Sirius, Vega, Procyon), we did not use the coronagraph because
it would make the acquisition process too long for a snapshot program
and its peripheral position in the detector would severely limit our
search radius. Placing the primary target behind the coronagraph would
also add uncertainty to the measurement of the position angle and
separation of any binary systems. Even with saturated central targets,
we could still search for companions, albeit with a lower sensitivity
closer to the central target (Table 2).

We coadded two sets of exposures for each target, resulting in a total
exposure time of 64s for the {\it F110W} and {\it F180M} filters and
128s for the {\it F207M} and {\it F222M} filters. Saturation of bright
targets and cosmic ray hits were minimized by using {\it NICMOS's}
multi-accumulate ({\it MULTIACCUM}) mode, which reads the detector in
a non-destructive manner at predetermined time intervals. In the event
of saturation or a cosmic ray hit, the {\it NICMOS} pipeline scales
the value from unaffected readouts so as to obtain the approximate
value due to the astronomical source. Only pixels that saturate or are
hit by a cosmic ray before the first readout at 0.303 s are lost. For
targets that were bright enough to saturate during the first readout,
we obtained photometry by using PSF fits.

Because there is no background
atmospheric glow, the extended PSF of the primary target is the
dominant source of background flux obscuring any fainter objects in
the field of view. We subtracted a properly scaled PSF of another star
of similar spectral type and brightness from the survey from the PSF
of each target.  A detailed discussion of the PSF subtraction process
is given in \citet[hereafter K98]{Kristetal1998}.  The quality of the
subtraction varied from target to target and depends primarily on
whether or not a good PSF match could be obtained. The PSF varies with
target color, telescope focus, and the position of the {\it NICMOS}
cold mask (K98). We were always able to find an isolated star whose
PSF was used as the reference for PSF subtraction. If the PSF
reference had been a close binary or if it had been contaminated by
background sources, we would have noticed a physically unrealistic
negative PSF in the subtracted image. We then performed aperture
photometry on the primary target as well as any other sources in the
field of view using standard IRAF routines and the aperture
corrections for encircled energy fraction listed in Table 2 of K98. To
verify the validity of the aperture corrections, we performed the
photometry of the crowded field of LHS 288 (31 sources, Figure 4),
varying the aperture from three to six pixels (0\farcs23 to
0\farcs46). The photometry agreed to $\lesssim$0.03 magnitudes in all
bands, regardless of aperture. For the final photometry we chose a six
pixel aperture, except in cases when a crowded field or a source near
the edge of the field required a smaller aperture.

\section{Determining the Sensitivity of the Search}

We define the ``sensitivity'' of the search as the extent to which we
can detect or rule out the existence of a companion to a given star at
a given separation and image contrast $\Delta$m.  The sensitivity
varies from target to target and is influenced by the overall
brightness of the primary target, the quality of the PSF subtraction,
the image filter, intrinsic detector noise, and the prominence of
detector artifacts. For each image these factors interact in a complex
way, thus making it difficult to draw generalizations about
instrumental sensitivity for the survey as a whole. We have therefore
devised a method to measure the sensitivity achieved for each target
at various separations, and quote individual results in Table 2.

Because {\it HST} is not subject to atmospheric effects, its images
are inherently stable, thus facilitating PSF modeling. We used {\it
Tiny Tim 6.3} \citep{KristandHook1997} to simulate generic {\it NIC2}
stellar PSFs through the four filters used in the search. The properly
scaled model PSFs were inserted into the PSF subtracted images of the
primary targets to test our ability to detect companions at a range of
contrasts and separations using a customized {\it IDL} code. At
sub-arcsecond separations, we inserted a single companion at
separations of 0\farcs2, 0\farcs4, 0\farcs6, and 0\farcs8 and a
varying range of contrasts at random position angles (Figure 5a).  The
PSF insertion code automatically excluded the strong diffraction
spikes present in well-exposed {\it NICMOS} images at 45\degr,
135\degr, 225\degr, and 315\degr. The residual flux from the PSF
subtracted primary target was set to zero at a radius interior to the
position of the artificially inserted companion to facilitate visual
inspection.  At separations of 1\farcs0 or greater, we produced an
image where artificial companions were arranged in a radial pattern
around the PSF subtracted primary (Figure 5b). This pattern tested the
sensitivity at separations of 1\farcs0, 2\farcs0, 3\farcs0, and
4\farcs0 at contrasts typically incremented from 6 to 13
magnitudes. The simulated images and their surface plots were then
visually inspected. In both regimes, an artificially inserted
companion was considered detectable if it was visible in the simulated
image and if a surface plot around the companion indicated that the
artificial PSF retained its characteristic stellar shape, with its
peak clearly above the background noise, corresponding to a typical
signal-to-noise of 3$-$5.  Although automating the PSF recovery
process (e.g., by using a cross-correlation algorithm) would have
saved a considerable amount of time, we were not convinced that
automated methods would appropriately distinguish between real
astronomical sources and residuals of the central star's PSF
subtraction, which can at times mimic star-like profiles.

\section{Results}

Other than the five companions reported in G04, which focused on
individual discoveries, we detected no further new companions during
this second phase of our search. We now report on the photometry,
astrometry, and search sensitivities attained during the survey. With
a statistically robust sample of 255 stars surveyed, these results
allow us to make assessments of the multiplicity of stars in the solar
neighborhood from a stellar population perspective.

\subsection{Color-Magnitude Diagrams}

We constructed color-magnitude diagrams for the twenty-four different
color-magnitude combinations from our observations through the four
filters. Because our sample includes only four certain substellar
objects (GJ 1001 B and C, GJ 229B, and 2MASSI J0559191- 140448), we
used synthetic photometry from the spectra of known L and T dwarfs to
better determine the form of the substellar sequence in this color
space. These values were obtained using flux-calibrated, near infrared
spectra \citep{Geballeetal2002,Knappetal2004}, weighted mean
trigonometric parallaxes \citep[and references
therein]{Golimowskietal2004b}, and the {\it NICMOS} Exposure Time
Calculator produced by STScI.  Figure 3 shows four color-magnitude
diagrams that are particularly well suited for mapping the stellar and
substellar main sequence. Main sequence targets and the
thirteen white dwarfs in the survey are labeled with large dots.
In these diagrams, we initially assume that
any object in the field of view of a primary target is a companion and
therefore shares the primary's trigonometric parallax. If the
assumption is correct, the object will fall within the stellar or
substellar sequence. Background objects, labeled with small dots,
appear as having unrealistically faint absolute magnitudes and tend to
cluster at the bottom of the diagrams. 

The trends in the {\it F110W$-$F180M} and the {\it F110W$-$F222W}
colors (Figures 3a and 3b) clearly indicate that the onset of CH$_4$
absorption happens sharply around the L6 spectral type, where the
colors turn blue. Although any single diagram may show an overlap
between the substellar sequence and the brighter background objects,
the degeneracy is broken when we consider that L and T dwarfs follow
different trends from the background sources in different color
combinations. The most dramatic example of these shifts appears in
Figures 3c and 3d, where methane imaging causes a large shift from
blue to red for the T dwarfs while the background sources show little
change.

\subsubsection{ Benchmark Objects}

GJ 1245ABC (labels 1, 2, and 4 in Figure 3) is an interesting system
containing three low mass components. In particular, GJ 1245C (4) is
one of the latest M dwarfs for which a dynamical mass is known. With a
mass of 0.074$\pm$0.013$M_\sun$ \citep{Henryetal1999}, this object
lies close to the theoretical hydrogen burning mass limit.

G 239-25B (label 3) was discovered during the first phase of this
search, and the implications of the multiplicity of the G239-25 system
are discussed in G04.  \citet{Forveilleetal2004} assign it a spectral
type of L0$\pm$1 based on near infrared spectra. This spectral
classification makes G 239-25B an important benchmark of the M/L
transition at the bottom of the main sequence. Its proximity in color
space to GJ 1245C re-enforces the importance of both objects as
benchmarks.

GJ 1001BC (labels 5, 6, and 7) was resolved as a double L4.5 dwarf,
and is discussed in detail in G04. The components of the system are
nearly equal in luminosity, and we plot them both individually (6 and
7) and combined (5) to illustrate how an equal flux binary appears in
the sequence. When compared to the L/T sequence outlined by the
synthetic photometry, both components of GJ 1001BC lie just before the
strong shift towards the blue that happens as a result of the onset of
CH$_4$ absorption. Their positions at this turning point are most easily
seen in the {\it F110W$-$F222M} color (Figure 3b). We are currently
working to refine the parallax of GJ 1001ABC, and to obtain dynamical
masses for the BC pair.

Finally, 2MASSI J0559191-140448 (T4.5, label 8) and GJ 229 B (T6,
label 9) are the only T dwarfs imaged in the survey, and serve as
confirmations that the sequence outlined by the synthetic photometry
agrees with real photometry. Whereas GJ 229B is a companion to the
M0.5V dwarf GJ 229A, 2MASSI J0559191-140448 is an isolated brown
dwarf. Its positions in panels a and b of Figure 3 illustrate how a
mid T dwarf can easily be mistaken for a white dwarf when more color
combinations are not used to break the degeneracy.

\subsubsection{Background Objects with Companion-Like Colors}

Figure 3 shows that there are several sources having colors that mimic
the colors of substellar companions in one or more panels. The
ambiguity is often accentuated when analyzing data sets with simpler
color combinations that were not designed a priori to discriminate
substellar objects (e.g.  {\it 2MASS JHK$_s$}). Interstellar reddening
considerations are particularly useful in identifying false
companions. Because the distance horizon of our search is only
$\sim$10 pc, any bona fide companions should not have appreciable
reddening in the near infrared. Conversely, distant main sequence or
giant stars may have significant reddening in the {\it F110W$-$F180M},
{\it F110W$-$F207M}, and {\it F110W$-$F222M} colors, which may place
background objects in the color space occupied by L and T dwarfs.  The
degeneracy is broken when considering the {\it F207M$-$F222M} and
especially the {\it F180M$-$F207M} colors, where the narrow spectral
coverage reduces the reddening of distant main sequence sources
(Figures 3c and 3d). Table 3 lists cases where the distinction between
a background object and a putative companion was particularly subtle
based on colors alone.  The white dwarfs as a group mimic late L and
early T dwarfs in {\it F110W$-$F180M} and {\it F207M$-$F222M}, but the
degeneracy is broken in {\it F180M$-$F207M}.

\subsection{Astrometry of Known Binaries}

High resolution images of nearby binary systems present opportunities
to map relatively short period orbits and therefore determine
dynamical masses. While actual orbital mapping is beyond the scope of
this work, we report the astrometry for select systems in Table 4. In
order to be listed in Table 4, both components of the system must have
been imaged simultaneously in the same {\it NIC2} field of view, and
the centroids must be determined to a precision better than $\pm$1
pixel.  The values we report are the weighted averages of separations
and position angles measured from the PSF centroids in all filters for
which saturation did not prevent reliable centroiding.  In the
simplest case of non-saturated and non-overlapping PSF cores, we adopt
a centroiding error of $\pm$0.1 pixel (G04).  Twelve out of the 19
pairs listed in Table 4 meet these criteria. The other seven pairs are
either very closely separated stars for which the PSF cores overlap
significantly or have central pixel saturation.  In either case, the
centroiding was determined using PSF fits. With the exception of the M
dwarfs, the majority of binaries in our survey had their PSF cores
saturated beyond the point where we could compute meaningful
astrometry. The precise value of the {\it NICMOS} plate scale varied
during HST cycle 7 (1997$-$1998) due to cryogen expansion that
distorted the dewar housing the detectors. 
To calibrate the plate scales for our observations, we used the values
tabulated by the Space Telescope Science Institute based on routine
monitoring of crowded star fields. For separations, the errors listed
in Table 4 take into account the four centroiding uncertainties ($x_a,
y_a, x_b, y_b$) added in quadrature.  For position angles, the errors
take into account the propagated centroiding errors.

\subsection{Results from the Sensitivity Search}

Table 2 lists the faintest detectable absolute magnitudes for putative
companions at a range of angular separations from each target star in
the survey. The distances and spectral types listed are based on the
best trigonometric parallaxes and spectral type estimates available in
the literature or unpublished trigonometric parallaxes recently
measured or improved by our group. It is important to note that each
line in Table 2 shows the results of one PSF insertion simulation, and
does not necessarily correspond to a single star. As described in the
notes column, a single PSF insertion simulation may have been done
around the two components of a resolved system if their separation was
small or if their contrast was large enough for the primary to
dominate the field. Because of these situations, the number of entries
in Table 2 is not meant to add up to the sample counts in Table 1. The
reader is referred to Table 1 for overall statistics of the sample and
to Table 2 for data on individual targets.

All targets were inspected for real companions visually in all four
bands over the entire field of view. Several factors must be
considered when choosing the best filter for the PSF insertion
simulations. Out of the four filters used in the search, the {\it
F110W} and {\it F180M} filters are the most suitable for close
separations ($\lesssim$ 0\farcs4) due to their narrower PSFs when
compared to the {\it F207M} and {\it F222M} filters.  Whereas L dwarfs
are brighter in {\it F180M} than in {\it F110W}, the T dwarfs are much
fainter in {\it F180M} due to methane absorption.  Although the {\it
F110W} band produces the narrowest PSFs due to its shorter wavelength
and is an intrinsically bright band for T dwarfs, we chose to report
the sensitivities in the {\it F180M} band for two reasons. First, our
uniform exposure time scheme ($\S$3) causes brighter targets to
saturate out to several pixels in the {\it F110W} band even in 0.303s,
decreasing our ability to probe the smallest separations. Second, the
width of the {\it F110W} PSF is comparable to the {\it NIC2} pixel
scale, causing a sharp spike on the central pixel (K98, Table 2). In
low signal-to-noise situations it becomes difficult to distinguish the
{\it F110W} PSF from bad pixels or other sharp artifacts introduced
during the PSF subtraction process. As discussed in $\S$5.4.2 our
sensitivity limit falls mostly in the L dwarf regime for sub-arcsecond
separations, and in the T dwarf regime for wider separations.  Based
on comparisons in particularly clear images, we estimate that using
{\it F110W} instead of {\it F180M} would increase our sensitivity by
$\sim$1 magnitude, but would pose an unacceptable risk of false
detections at close separations. We therefore uniformly report
sensitivities for all separations in {\it F180M}, but emphasize that
those values can be safely transformed to {\it F110W} limiting
magnitudes for separation greater than 1\farcs0 by adding 1.0
magnitude to the {\it F180M} limits in Table 2. Because late T dwarfs
appear the faintest in {\it F180M}, a detection in that band also
implies detection in the other three bands, therefore providing the
color information needed to characterize the object. Listing our
simulation results in the {\it F180M} band therefore maximizes the
instrumental dynamic range of the images while still providing the
sensitivity needed to characterize T dwarfs.

\subsection{The M Dwarfs}

Of the 188 star systems imaged within 10 pc in this survey, 126
systems have M dwarfs as the primary (or single) component\footnote{GJ
169.1AB is an M4.0V/white dwarf binary. Although the brighter M4.0V
component is generally considered to be the primary component, the
current situation does not reflect the components' masses or spectral
types at the time of stellar formation and main sequence evolution,
when the current white dwarf was much more massive and luminous than
the M dwarf. We therefore do not consider GJ 169.1AB to be a system
with an M dwarf primary.}.  Because these M dwarfs were selected
randomly from a volume limited sample based on their trigonometric
parallaxes, the sub-sample lends itself well to statistical
considerations.  We now apply the sensitivity limits in Table 2 to
derive the multiplicity fraction for M dwarfs under several scenarios.

\subsubsection{Establishing Search Completeness for M Dwarfs}

Figures 6 a and b show the ranges in sensitivities obtained for M
dwarfs at each of the eight angular separations probed by the PSF
insertion simulations. In Figure 6b, we used the known distance to
each target to convert contrasts into absolute magnitudes, and relate
these absolute magnitudes to the spectral types of putative
companions. Because sensitivity is a complex function of contrast,
instrumental background, apparent magnitude, and the quality of the
PSF subtraction, there is a significant spread about the mean values
quoted in Figure 6. Overall, we would detect companions with
$\Delta${\it F180M}$=$2.5 to 10.2 magnitudes at separations of
0\farcs2 to 4\farcs0, respectively.

In order to transform our observational sensitivities (Figure 6a) to
astrophysical parameters, we substitute physical separations in AU in
place of angular separations and apply the statistical relation
between physical separation and semi-major axis for a sample of
binaries with random inclinations and eccentricities,$<a> =
1.26<\rho>$ \citep{FisherandMarcy1992}, obtaining Figure 7.  The large
plus signs in Figure 7 indicate the 90\% detection limits for
semi-major axes ranging from 0 to 40 AU, binned in 2 AU increments. We
assume a flat contrast curve for sensitivities beyond 40 AU.  Because
of the large factor in distance covered by this volume-limited search,
the 90\% detection limits in physical separation are effectively
established by the most distant stars in the sample. It is possible to
boost sensitivity at closer physical separations by establishing a
closer distance horizon for the search, at the expense of overall
sample size. We examined the effect of using a closer distance horizon
for calculating sensitivity limits, and came to the conclusion that it
is more important to maintain a robust sample, especially because more
sensitive but much smaller studies have already been done
\citep[e.g.,][]{Closeetal2003}. We emphasize that Table 2 contains all
data necessary for different statistical formulations, and is
available in machine readable format in the online version of this
paper.

We also considered the effect that the small field of view of {\it
NIC2} (19\farcs5$\times$19\farcs5) has on sample completeness at large
physical separations. Figure 8 is a histogram displaying the number of
all M dwarfs, including resolved system secondaries, sampled within 10
pc (N$=$141) as a function of outer search radius, binned in 10 AU
increments. While all M dwarfs were probed to semi-major axes as close
as 5 AU\footnote{As noted in Table 2, GJ 15A, LHS 224AB, GJ 623AB, and
GJ 644ABD are M dwarfs for which core saturation prevented the
establishment of a sensitivity limit at 0\farcs2. A single M dwarf
system, GJ 747AB, saturated out to 0\farcs4. All of these cases
correspond to statistically corrected semi-major axes smaller than 5
AU.}, only the farthest 12 targets were probed at semi-major axes
greater than 120 AU. In order to retain the statistical significance
of our sample, we consider only physical separations corresponding to
mean semi-major axes between 5 and 70 AU (100\%  to 79.4\% complete),
and divide the number of companions found in the bins from 40 to 70 AU
by that bin's completeness fraction.

\subsubsection{The M Dwarf Multiplicity Fraction}

Table 5 lists companions to M dwarfs in our sample within our
completeness range of 5 to 70 AU that were re-detected in our search
or are new companions discovered during this search and published in
G04. Combining these known binaries to the null detections and
sensitivity limits we present in Figures 6$-$8, we now present formal
multiplicity fractions for three distinct combinations of companion
types and ranges in semi-major axes. These results are summarized in
Table 6. In each case, the 1$\sigma$ confidence intervals were
calculated using the binomial distribution approach outlined by
\citet{Burgasseretal2003}. This approach is preferable to traditional
Poisson statistics whenever the probability distribution is
non-Gaussian. In each of our three different scenarios discussed below
the multiplicity fraction is low enough that even with the sample
of 126 systems, the probability distribution is not
symmetric about the central peak value because proximity to the
limiting case of a multiplicity fraction of zero causes a sharper
drop-off towards the lower limit of the probability distribution
(Figure 9). Given a multiplicity fraction $\epsilon_m$, the
probability distribution of finding $n$ binaries in a sample of $N$
systems is governed by
$$P(n)=\frac{N!}{n!(N-n)!}\epsilon_m^n(1-\epsilon_m)^{N-n}.$$ This
relationship can be inverted to solve for the probability distribution
of a given multiplicity fraction given the observational results $N$
and $n$, yielding $$P'(\epsilon_m)=(N+1)P(n),$$ which can then be
integrated numerically to find the lower and upper limits of
$\epsilon_m$ corresponding to 68\% (1$\sigma$ for a Gaussian
distribution) of the area under the probability distribution curve, as
shown by the shaded areas in Figure 9.

\subsubsubsection{The M Dwarf Multiplicity Fraction for M0V to M9V Companions at Separations of 5 to 70 AU}

At an inner search radius of 5 AU, our search is 90\% sensitive to
$M_{F180M} \lesssim 11.2$, corresponding to early L spectral types
(Figure 7a). Eleven of the 12 known companions listed in Table 5 are M
dwarfs meeting this sensitivity criterion. Five of these companions
lie between 40 and 70 AU, where the completeness of the search is
reduced due to the limited field of view. Placing these five systems
into the separation bins shown in Figure 8 yields 2 systems in the
40-50 AU bin, 2 systems in the 50-60 AU bin, and one additional system
in the 60-70 AU bin. Dividing these numbers by the fractional
completeness of these bins (0.957, 0.879, and 0.794) and summing the
results yields 5.62. We then transform the multiplicity obtained at
90\% confidence level to a true volume limited multiplicity fraction
by dividing 11.62 (the sum of 5.62 and the remaining companions from 5-40 AU)

by 0.9, obtaining 12.91.  Rounding this number up to
13, we see that we would likely have recovered 2 additional real
companions with separations ranging from 5 to 70 AU.  Applying the
binomial distribution, we conclude that the multiplicity fraction for
M dwarf companions orbiting M dwarf primaries at semi-major axes from
5 to 70 AU is $\epsilon_m=10.3_{-2.1}^{+3.4}\%$ (Figure 9a).

\subsubsubsection{The M Dwarf Multiplicity Fraction for L0 to L9
Companions at Separations of 5 to 70 AU}

Although our search did not detect any L dwarf companions within 10 pc
and in the separation regime of 5 to 70 AU,\footnote{GJ 1001 B and C
are beyond 10 pc \citep{Henryetal2006}.}  it is possible to assign a
multiplicity fraction based on completeness arguments.  Figure 7a
shows that at 5 AU, the detection rate for L dwarfs is only
$\sim$50\%.  It is not possible to obtain a truly volume
limited multiplicity fraction in this separation range. We therefore
constrain the sample to include only the 51 systems for which the
detection of an L9 companion at 5 AU is possible.  Applying the
binomial distribution, we obtain a multiplicity fraction of of
$\epsilon_m = 0.0_{-0.0}^{+3.5}\%$ (Figure 9b). An alternative
approach is to maintain the volume limited nature of the sample by
increasing the inner limit of the separation range. From Figure 7a,
the inner radius at which $>$90\% of the systems were probed is 12 AU. We
therefore calculate a volume limited multiplicity fraction for L0 to
L9 companions to M dwarfs of $\epsilon_m = 0.0_{-0.0}^{+1.4}\%$ valid
at separations ranging from 12 to 70 AU.

\subsubsubsection{The M Dwarf Multiplicity Fraction for L0 to T9 Companions from 10 to 70 AU}

Our sensitivity to T dwarfs at close separations is diminished due to
their intrinsic faintness.  We therefore restrict the inner search
radius to 10 AU, where the search was 90\% sensitive to L dwarfs and
$\sim$50\% sensitive to late T dwarfs. At separations beyond 12 AU,
Figure 7b indicates considerable scatter in the 90\% sensitivity
limits. Based on the trend on Figure 7b, we adopt a 90\% sensitivity
limit of {\it M$_{F110W}$}$=$ 17.5, corresponding to spectral type
$\sim$T9. One T6 dwarf, the class prototype GJ 229B, was detected at
an inferred semi-major axis of 55.3 AU. Following the same approach we
used for the L dwarfs, we calculate the multiplicity fraction for a
sub sample as well as for the volume limited sample. There were 43
systems for which a late T dwarf detection at 10 AU was possible. This
sub-sample yields a multiplicity fraction of $\epsilon_m =
2.3_{-0.7}^{+5.0}\%$ The complete sample is sensitive to late T dwarfs
at separations $\geqq$ 14 AU. We therefore calculate a volume limited
multiplicity fraction of $\epsilon_m = 0.8_{-0.2}^{+1.8}\%$ valid at
separations ranging from 14 to 70 AU.

\section{Discussion}

\subsection{Sensitivity to Companion Masses}

Estimating masses for field brown dwarfs is a difficult
problem. Whereas the masses of main sequence stars can be estimated
from mass-luminosity relations
\citep{HenryandMcCarthy1993,Henryetal1999,Delfosseetal2000}, brown
dwarfs are constantly cooling, and therefore have a
mass-luminosity-age relation. Such a relation has not yet been
established empirically. Currently, the best way of estimating brown
dwarf masses is by correlating spectral types to effective
temperatures, and then checking the effective temperature against
evolutionary model predictions, assuming a certain age for the brown
dwarf in question.  This approach is heavily model dependent, and the
end result of such calculation can at best serve as a guideline for
the mass range for a particular object. With this caveat in mind, we
now apply this approach to the limiting spectral types we report in
Table 6.

Assuming a mean age of 3 Gyr for the nearby L dwarf field population
\citep{Seifahrtetal2010}, the effective temperatures for brown dwarfs
of spectral types L3, L5, L8, T5, and T7 are estimated to be 2000K,
1750K, 1500K, 1200K, and 900K, respectively
\citep{Golimowskietal2004a,Cushingetal2008}. Adopting the evolutionary
models of \citet{Chabrieretal2000}, we estimate
approximate masses of 0.073 M$_{\sun}$, 0.070 M$_{\sun}$, 0.057
M$_{\sun}$, 0.052 M$_{\sun}$, and 0.040 M$_{\sun}$ for spectral types
L3, L5, L8, T5, and T7 (Table 6). The last number has considerable
uncertainty due to the steeper decline in effective temperatures for
subtypes later than $\sim$T5 and the need to extrapolate the Chabrier
models at low temperatures. We therefore adopt 0.040M$_{\sun}$ at 3
Gyr as a guideline for the minimum mass detectable by our search. We
note that the scatter in age in the nearby field population is likely
to cause a large dispersion in the masses of detectable
objects. Unless there are further data indicative of the age of an
individual brown dwarf, the mean value we adopt here should be used
with extreme caution.

\subsection{A Current Map of the Brown Dwarf Desert}

The idea of the brown dwarf desert continues to evolve. The term was
originally used to describe the fact that radial velocity 
surveys of solar analogs detect an abundance of extra-solar
planets but rarely detect brown dwarfs, even though a brown dwarf's
higher mass makes its detection easier. In their seminal work,
\citet{MarcyandButler2000} found that $<$1\% of main sequence Sun-like
stars harbor brown dwarfs. Several other studies have since then
obtained similar results for different ranges in separation, primary
mass, and system age. \citet{Oppenheimeretal2001} conducted the first
successful search for brown dwarf companions, discovering the T dwarf
prototype GJ 229B. Their infrared coronagraphic search of stars within
8 pc detected a single substellar object, from which they cautiously
imply a stellar/substellar multiplicity fraction of $\sim$1\%.
\citet{McCarthyandZuckerman2004} used Keck coronagraphy to search 102
nearby field GKM stars at separations from 75 to 1200 AU. They found
one brown dwarf companion, and report a binary fraction of
1$\pm$1\%\footnote{Using the binomial distribution treatment we adopt
in this paper, 1 detection out of 102 observations is equivalent to a
multiplicity fraction of 1$^{+3}_{-0.2}$\%.}. We note that their
result agrees well within statistical uncertainties to our results
(Table 6), suggesting a wide desert with no significant change in the
substellar companion fraction from 10 to 1200
AU. \citet{Luhmanetal2005} used {\it HST's} Wide Field Planetary
Camera 2 ({\it WFPC2}) to survey 150 members of the young cluster IC
348 ($\sim$2 Myr) at separations of 120$-$1600 AU. Of these stars, 85
were in the mass range 0.08$-$0.5 M$_{\sun}$, approximately
corresponding to the mass range for main sequence M dwarfs
\citep{HenryandMcCarthy1993}. They found one possible substellar
companion to a low mass star, but note that it is not possible to
ascertain companionship due to the wide separation of this system
($\sim$1400 AU). Based on this finding, \citeauthor{Luhmanetal2005}
derive an upper limit of 4\% for the substellar companion fraction of
low mass stars.  This result is again in very good agreement with our
results, suggesting that there is little evolution in the multiplicity
fraction of low mass stars after the first few million years, and
again suggesting no significant change in the substellar companion
fraction beyond 10 AU.  Regarding single objects, Luhman et al. find
that 14 out of 150 objects are likely substellar based on evolutionary
models \citep{Chabrieretal2000}. They note that the fact that they
detect ten times more isolated stars than isolated brown dwarfs in IC
348 indicates that the brown dwarf desert may not be limited to the
formation of companions, but may also extend to the formation of
single objects. \citet{MetchevandHillenbrand2009} used adaptive optics
on Keck and Palomar to survey 266 Sun-like (F5$-$K5) stars, and infer
a brown dwarf companion frequency of
3.2$^{+3.1}_{-2.7}$\%~\footnote{2$\sigma$ limits.}  for separations of 28
to 1590 AU. Finally, direct imaging searches for planetary companions
would be capable of detecting brighter brown
dwarfs. \citet{Masciadrietal2005} used {\it VLT/NACO} to search 30
young ($<$200 Myr) GKM stars and found no brown dwarf or planetary
companions at separations larger than 36 AU. In a similar fashion,
\citet{Billeretal2007} used {\it VLT} and {\it MMT} to search 45 young
GKM field stars at separations of 20$-$40 AU, and also found no brown
dwarfs. Due to smaller sample sizes, the last two studies do not add
significant constraints to the brown dwarf desert, but their null
detections are certainly in agreement with constraints set by the
larger studies.

The sum of these studies, along with the results we present in this
paper, indicate a consistent image of a brown dwarf desert that is
mostly invariant with respect to the mass of the primary star, and
which is valid for a wide range of separations ranging from 5 AU to
1600 AU. Whether the search is sensitive to substellar companions to
Sun-like stars at intermediate to large separations
\citep{MetchevandHillenbrand2009}, substellar companions to low mass
stars at intermediate separations (our results), or a mixture of young
stars with masses ranging from solar down to the M dwarf regime
\citep{Masciadrietal2005,Billeretal2007} the detection rate is always
consistent with a stellar-substellar binary fraction on the order of a
few percent.

\subsection{Is the Desert Real?}

The multiplicity fraction of Sun-like stars is $\sim$50\%
\citep{DuquennoyandMayor1991,Raghavanetal2010}.  The multiplicity rate
for stellar companions to M dwarfs at all separations is $\sim$30\%
\citep{HenryandMcCarthy1990,Henry1991PhD,FisherandMarcy1992}. Based on
our results (Table 6) and the companion searches we discuss in
$\S$6.2, it is clear that stellar companions outnumber brown dwarf
companions by a factor $\gtrsim$10. Does this paucity of brown dwarfs,
however, constitute a ``real desert''? A few studies
\citep[e.g.,][]{MetchevandHillenbrand2009,GretherandLineweaver2006}
have suggested that the dearth of brown dwarf companions is a natural
consequence of a well behaved, Salpeter-like \citep{Salpeter1955}
universal Companion Mass Function (CMF) that tends to lower
multiplicities at lower mass ratios, and that a real brown dwarf
desert would only exist if the observed number of brown dwarf
companions is significantly lower than what a universal CMF would
predict. In particular, \citeauthor{GretherandLineweaver2006} note
that the overlap of the planetary CMF and the stellar CMF reaches a
minimum at $\sim$0.03 M$_{\sun}$, causing the observed paucity of
brown dwarf companions.  In our search, we test the hypothesis of a
universal CMF by focusing primarily on low mass stars. As shown in
Table 5, the twelve M dwarf binaries we detected between 5 and 70 AU
have primary masses ranging from $\sim$0.6 to $\sim$0.1
M$_{\sun}$. Figure 10 is a plot of the masses of the primary and the
secondary components of this sample. Figure 10 shows that the mass
ratios of low mass binaries tend to increase (i.e. approach equal mass
components) as masses approach the hydrogen burning limit, thus
excluding the formation of brown dwarf secondaries. Our completeness
analysis demonstrates that this trend is not an observational
selection effect.  Indeed, detecting companions with higher contrasts
is easier for intrinsically fainter primary stars, so the selection
effect works against the trend noted in Figure 10. Reconciling our
observations with the idea of a universal CMF would require this
function to be rather restricted in the sense that it would not be a
function of mass ratio, or would only be valid for Sun-like stars.
For any reasonably broad definition, we conclude that deviations from
a universal CMF do exist in the brown dwarf
regime. The brown dwarf desert is therefore a reality whether one
defines it in terms of total numbers or in terms of a deviation from a
trend. We advocate that the concept of a universal CMF
is probably not a useful representation of Nature.

\subsection{More Evidence for A Discontinuity at the Hydrogen Burning Limit?}

VLM binaries have a strong tendency towards high (i.e.~unity) mass
ratios \citep[e.g.][]{Burgasseretal2007}. The effect has been
demonstrated to be an intrinsic characteristic of VLM stars and brown
dwarfs through Bayesian analysis \citep{Allen2007}. Our results
(Figure 10) show that mass ratios tend to increase as stellar masses
approach the hydrogen burning limit, with the strong onset of nearly
equal mass duplicity happening somewhere between 0.2 and 0.1
M$_{\sun}$. Other studies have also suggested that the basic
population properties of Initial Mass Function (IMF), CMF, and the
binary separation distribution all appear to change significantly at a
mass of $\sim$0.1 M$_{\sun}$, slightly above the hydrogen burning
limit. \citet{Closeetal2003} conducted an adaptive optics search of 39
VLM objects with spectral types ranging from M8.0V to L0.5, and found
a mass distribution similar to the one shown in Figure 10 (see their
Table 3). They also probed smaller separations than our formal limit
of 5 AU, and found that whereas higher mass stars have a separation
distribution peaked at 30 AU \citep{DuquennoyandMayor1991}, VLM
binaries have a separation distribution peaked at 4 AU. Also, Bayesian
analysis of several studies \citep{Allen2007} demonstrates that VLM
and brown dwarf binaries with separations $>$ 20 AU are extremely
rare. We note that \citeauthor{Closeetal2003} probed significantly
smaller separations than we did, but did not establish formal
detection limits. \citet{Krausetal2005} conducted a search for VLM
binaries in the Upper Scorpius OB association, and also found results
consistent with a discontinuity in the separation distribution at a
mass of 0.1M$_{\sun}$.

In an analysis of data from several open cluster studies,
\citet{ThiesandKroupa2007} demonstrate that the observed mass
distribution is incompatible with the existence of an IMF that is
monotonic about the hydrogen burning limit. They note that because
stellar formation and stellar ignition are in principle unrelated
processes governed by different areas of physics, there is no reason
to expect that the IMF discontinuity would be caused by the onset of
hydrogen burning. They therefore allow for an arbitrary overlap of the
stellar and brown dwarf components of the IMF, thus allowing for a
smooth turnover. In light of our companion mass distribution for low
mass stars (Figure 10), new developments in the hydrodynamical
simulations of star cluster formation \citep{Bate2009a,Bate2011}, and
new observations of young stellar clusters
\citep{Krausetal2008,Krausetal2011,Evansetal2012}, we re-examine the
nature of the IMF discontinuity at masses close to the hydrogen
burning limit.

The details of the mass function for older field objects close to the
hydrogen burning limit are difficult to quantify. The difficulty is
mostly due to the lack of a robust volume limited census of L and T
dwarfs based on trigonometric parallaxes or reliable distance
estimates (errors $<$20\%). For the M dwarfs, the situation is more
clear. Our recent results from the {\it RECONS} 10 pc census indicate
a {\it minimum} M dwarf space density of $0.057pc^{-3}$
\citep{Henryetal2006}\footnote{$0.059pc^{-3}$ for epoch 2012.0. See
www.recons.org for the latest numbers and analysis. 
Comparison of the 10 pc sample with the 5 pc sample indicates that the 
10 pc M dwarf sample is $\sim$70\% complete. We note, however, that an
analysis of the {\it RECONS} sensitivity limits indicates that the  
assumption of a representative M dwarf sample within 5 pc may be 
significantly biased by statistics of small numbers.}.
\citet{Cruzetal2007} find a space density of $4.9 \times 10^{-3}
pc^{-3}$ for M dwarfs later than M7V and a lower limit of $3.8 \times
10^{-3} pc^{-3}$ for L dwarfs. Assuming that field age ($\sim$1$-$5
Gyr) brown dwarfs with masses slightly below the hydrogen burning
limit are predominately mid to late L dwarfs ($\S$6.1, Table 7), and
that stars of spectral type M7V or later have masses $\lesssim$0.1
M$_{\sun}$ \citep{Henryetal1999,Delfosseetal2000}, the ratio of
objects with masses above and below 0.1 M$_{\sun}$ is 6.9. The shape
of the M dwarf distribution in the {\it RECONS} 10 pc census
corresponds broadly to the distribution of our {\it NICMOS}
sample, (Figure 1), with the drop-off happening at around spectral
type M6V, corresponding to $\sim$0.1 M$_{\sun}$. Even if the actual
density for L dwarfs is a few times greater than the lower limit of
\citet{Cruzetal2007}, there is still a significant difference in the
number of stars versus brown dwarfs.

Could the onset of core hydrogen fusion cause the discontinuity in the
IMF and the CMF via a radiative feedback mechanism? We caution that
our understanding of stellar formation processes in this mass range is
rather limited from a theoretical as well as an empirical basis, so
any explanation is tentative at best. We speculate that if the onset
of core hydrogen burning at ages from 3$-$5 Myr
\citep{ChabrierandBaraffe1997} has a significant role in hindering
accretion, the star formation process would produce a discontinuity at
masses $\gtrsim$ 0.075 M$_{\sun}$. Although it has been generally
accepted that protostars acquire the bulk of their mass during the
first 1 Myr, observations show that a sizable fraction of substellar
objects continue to accrete for a much longer
time. \citet{Jayawardhanaetal2003} find that 40\%$-$60\% of brown
dwarfs in young star forming regions with ages up to $\sim$10 Myr show
infrared excesses consistent with accretion. There is also
observational evidence that at least some high mass brown dwarfs
undergo phases of strong accretion comparable to the T Tauri phases of
more massive stars \citep{Bouyetal2008,Comeronetal2010}.  For the
highest mass proto-brown dwarfs, late accretion may be enough to
ignite hydrogen fusion, or to otherwise significantly change the
manner in which the young object interacts with its environment. More
observations and theoretical work are needed to confirm or discard
this hypothesis, in particular with regards to testable predictions of
accretion rates. Even if late accretion brings the total mass of a
proto-brown dwarf above the hydrogen burning limit, we lack a clear
understanding of how the onset of core hydrogen fusion would hinder
accretion. At ages of a few Myr, the vast majority of an object's
luminosity comes from the release of internal gravitational energy, so
the onset of hydrogen burning would have a negligible effect on
overall luminosity \citep{ChabrierandBaraffe1997}.  We note, however,
that hydrodynamical cluster collapse simulations \citep{Bate2009a} are
in good agreement with the stellar IMF and stellar CMF, but
overproduce the number of brown dwarfs unless radiative feedback is
incorporated into the model \citep{Bate2011}. The last model produces
a cluster of stars and brown dwarfs whose statistical properties are
very similar to those of observed young clusters, suggesting that
radiative feedback is indeed an important mechanism in brown dwarf
formation. If the discontinuity in the CMF and the IMF at 0.1
M$_{\sun}$ stands up to further observational scrutiny, a strong
convergence of theoretical models and observational evidence will be
needed to prove this or other hypotheses.

\section{Conclusions}

We conducted a large, volume-limited, high resolution search for
substellar companions around nearby stars, with a particular emphasis
on M dwarf systems.  By evaluating the completeness and sensitivity of
the search, we have established the multiplicity fractions for M
dwarfs listed in Table 6. We find a multiplicity fraction of $0.0_{-0.0}^{+3.5}$\% 
for L companions to M dwarfs with semi-major axis
ranging from 5 to 70 AU.  Including T dwarfs down to spectral type T9
and restricting the inner search radius to 10 AU yields a multiplicity
fraction of $ 2.3_{-0.7}^{+5.0}$\%. These rates are far less than
M dwarf pairs, for which we found a multiplicity fraction of
$10.3_{-2.1}^{+3.4}$\% for separations of 5 to 70 AU. Based on these
results, we summarize the substellar multiplicity fraction for M
dwarfs as being on the order of a few percent or less.  As discussed
in $\S$6.2, several other multiplicity studies have reached
essentially the same conclusion regardless of primary mass, the
separations probed, or the sample's age estimate. The emerging picture
is that of a pervasive ``brown dwarf desert'', hinting to origins that
are largely independent of a binary's formation mechanism.  By
specifically focusing on low mass primaries, our study has weakened
the case for mass ratio dependence in the formation of substellar
companions. We add ours to a list of several studies that indicate
that the companion mass function is truncated at a mass $\sim$0.1
M$_{\sun}$, slightly above the hydrogen burning mass limit ($\S$6.4,
Figure 10). While the primary focus of this work is characterizing the
companion population, we also note in $\S$6.4 that the results we
obtain for the stellar/substellar multiplicity fraction are consistent
with estimates for the population density of isolated brown
dwarfs. The similarity suggests that mechanisms causing the observed
paucity of brown dwarfs, both as companions and as isolated objects,
may be intrinsic to the brown dwarf formation process. Recent results
from hydrodynamic cluster collapse simulations as well as evidence for
T Tauri like accretion at ages of a few Myr make radiative feedback
from recently ignited very low mass stars a good candidate mechanism
for truncating the IMF and the CMF at masses slightly above the
hydrogen burning mass limit.

The authors thank Russel White, Deepak Raghavan, and Adam Burgasser
for useful conversations. We thank our anonymous referee for many insightful
comments that greatly improved the paper. This work is based on observations
made with the NASA/ESA
{\it Hubble Space Telescope}, obtained at the Space Telescope Science
Institute, which is operated by the Association of Universities for
Research in Astronomy, Inc., under NASA contract NAS 5-26555. These
observations are associated with {\it HST} programs 7420, 7894, and
9485. Support for these observations was provided by NASA through
grants HST-GO-07420, HST-GO-07894, and HST-GO-09485 from the Space
Telescope Science Institute (STScI).


\bibliographystyle{apj}
\bibliography{nicmosreferences}


Viana, A., Wiklind, T., et al. 2009, NICMOS Instrument Handbook, Version 11.0, (Baltimore: STScI).

Thatte, D. and Dahlen, T. et al. 2009, NICMOS Data Handbook, version 8.0, (Baltimore, STScI)


\begin{figure}
 \begin{center}
 \includegraphics[scale=0.7,angle=90]{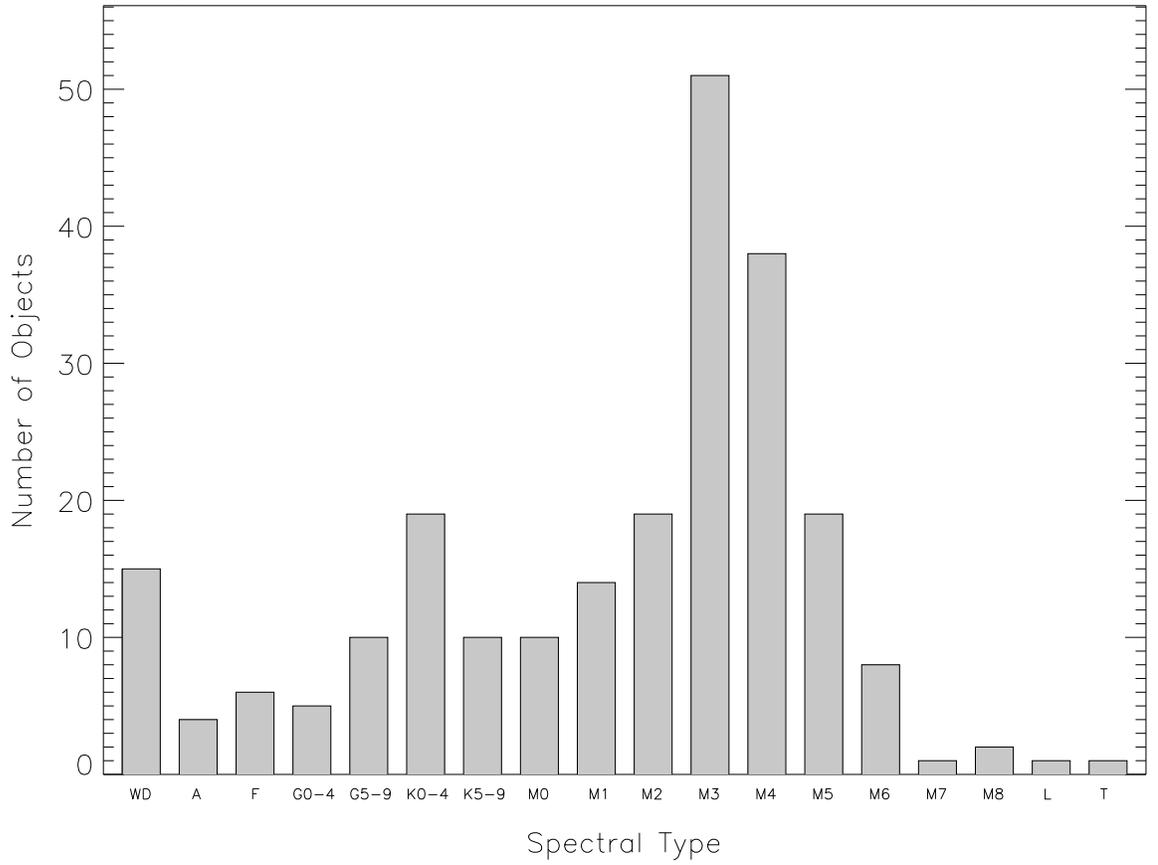}
\figcaption[figure1] {Spectral type distribution of the 239 
targets within 10 pc in the search. The sample constitutes 69\% of the {\it RECONS}
10 pc sample (epoch 2012.0). 63\% of the targets are M dwarfs, which is in 
close agreement with the M dwarf distribution of the 10 pc sample, 69\% (epoch 2012.0). 
\label{figure1}}
\end{center}
\end{figure}

\begin{figure}
 \begin{center}
 \includegraphics[scale=0.7,angle=90]{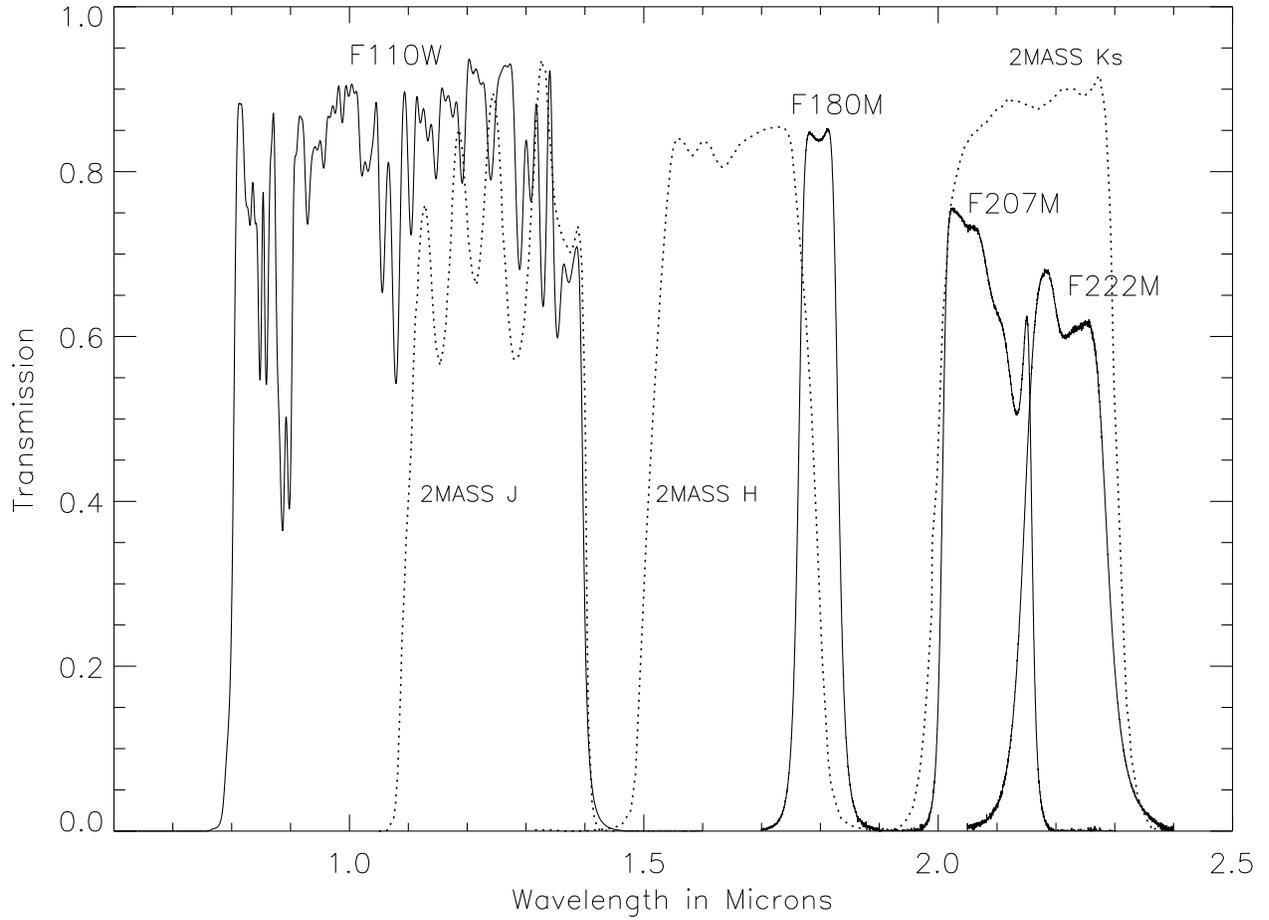}
\figcaption[figure2] {Transmission curves for the four {\it NICMOS}
filters used in the survey. The {\it 2MASS} filters are plotted with dotted
lines for comparison. Although no individual {\it NICMOS} filter is a close
match to a ground based equivalent, together they cover nearly the same wavelength
range of ground based near infrared color systems.}

\end{center}
\end{figure}




\begin{figure}[ht]
 \begin{center}
    \subfigure[]
	{\includegraphics[scale=0.37,angle=0]{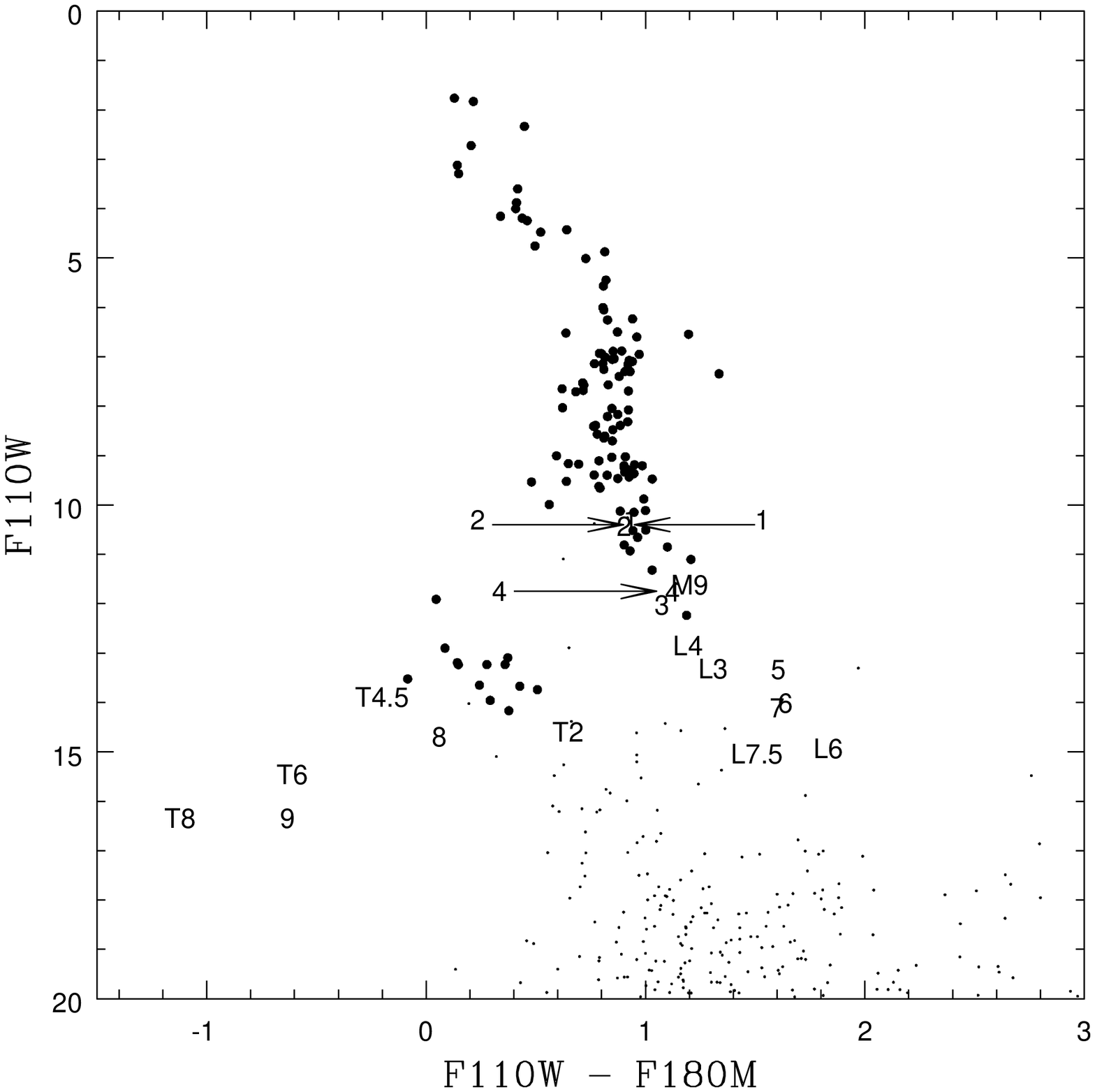}}
    \subfigure[]
	{\includegraphics[scale=0.37,angle=0 ]{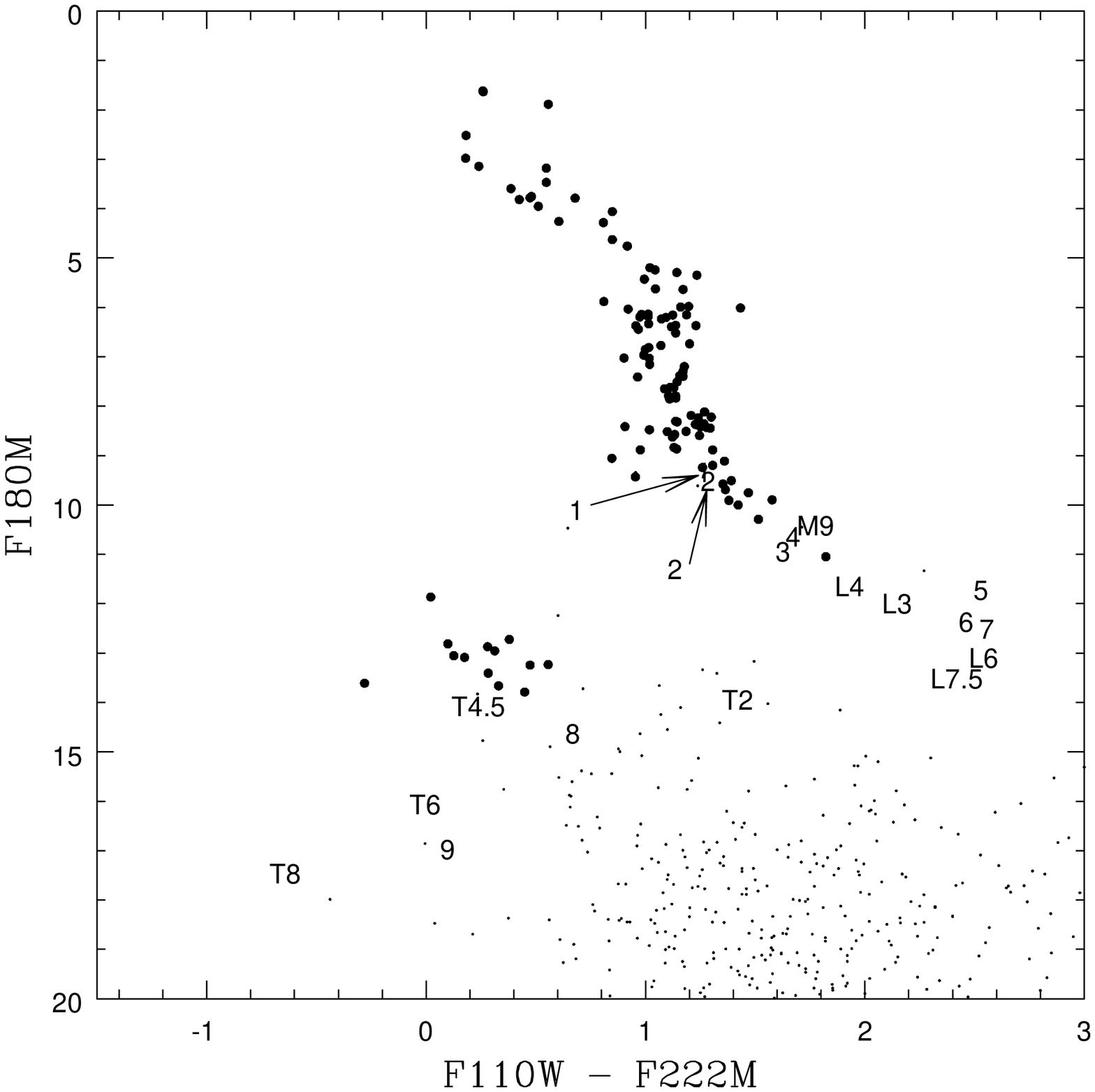}}
    \subfigure[]			    
	{\includegraphics[scale=0.37,angle=0 ]{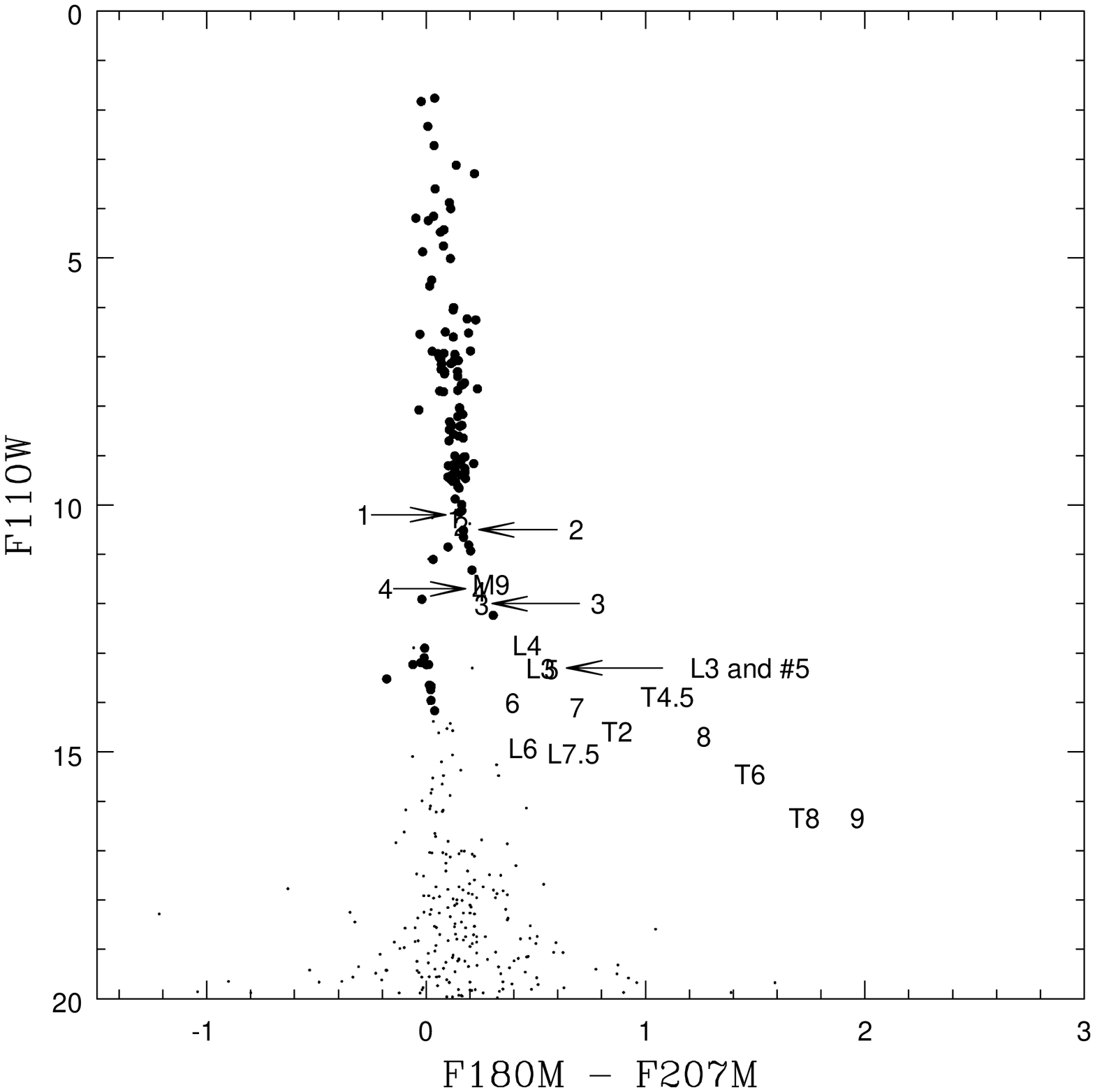}}
    \subfigure[]			    
	{\includegraphics[scale=0.37,angle=0 ]{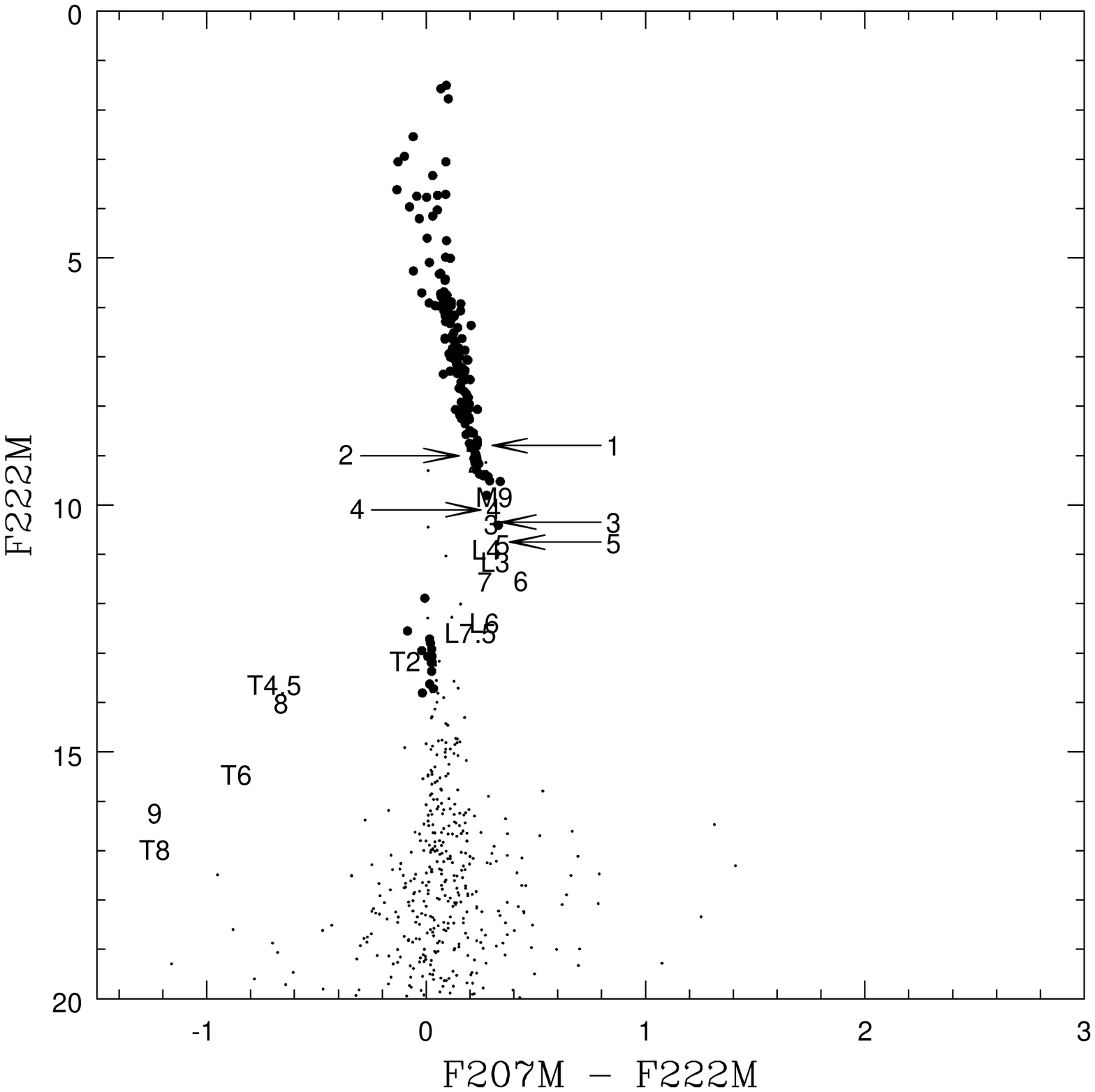}}
 \figcaption[figure3] { Selected color-magnitude diagrams designed to
 detect substellar companions. The large dots are the primary targets
 of the search, including 13 white dwarfs.  The small dots are background objects. Synthetic
 photometry of L and T dwarfs, as well as one M9 dwarf, is plotted
 using a label for spectral type, with the precise dot position at the
 center of the label. In these diagrams, all objects within the field
 of view of a primary target are plotted assuming a common parallax
 (i.e. companionship). Only if the assumption is correct would the
 object fit in the stellar or substellar sequence.  The
 benchmark objects discussed in $\S$5.1.1 are labeled as follows: (1) GJ
 1245A, (2) GJ 1245B, (3) G 239-25B, (4) GJ 1245C, (5) GJ
 1001BC (combined), (6) GJ 1001B, (7) GJ 1001C, (8) 2MA 0559-1404, and (9)
 GJ 229B. Panels (a) and (b) illustrate the drastic color shift around
 spectral type L6 caused by the onset of CH$_4$ absorption.  The
 reduced effect of interstellar reddening on background objects
 displayed in panels (c) and (d), as well as the large shift from red
 to blue for substellar objects, make these bands particularly useful
 for methane imaging.}

\end{center}
\end{figure}

\begin{figure}[ht]
 \begin{center}
    \subfigure[]
	{\includegraphics[scale=0.65, angle=0, clip=, bb= 92 92  495 495]{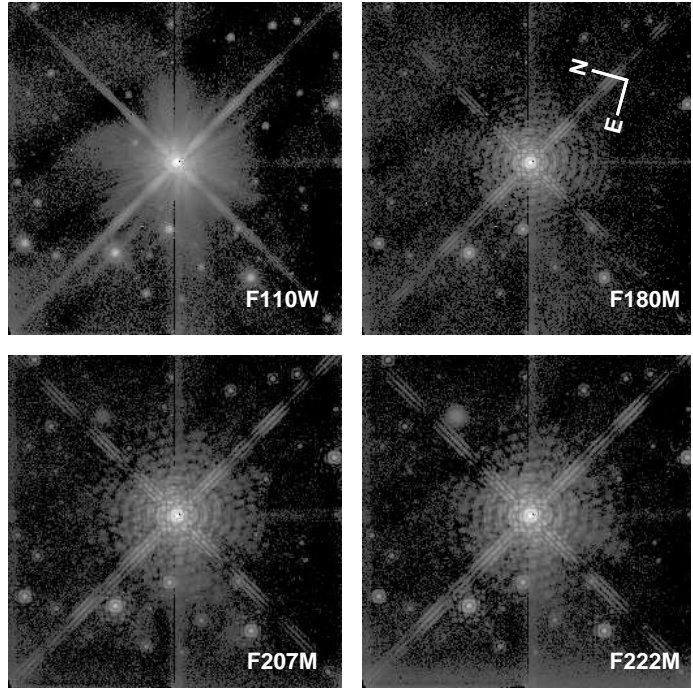}}
    \subfigure[]
	{\includegraphics[scale=0.65, angle=0, clip=, bb= 92 92 495 495]{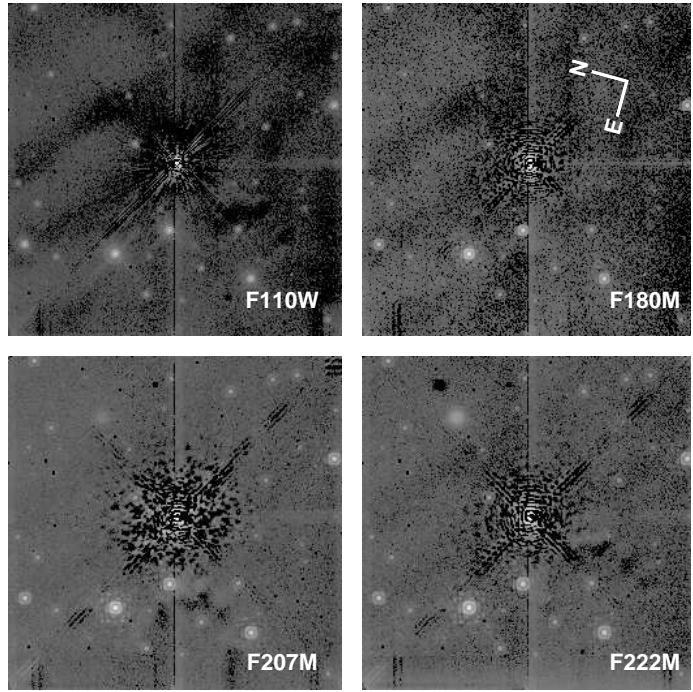}}
    \figcaption[figure4] { Survey images for LHS 288 (M5.5V) using
logarithmic scaling.  The frames illustrate typical survey images both
before (a) and after (b) PSF subtraction.  The ghost-like
coronagraphic artifact is visible in the upper left hand corner,
particularly in the {\it F222M} images.  The highly structured PSF of
the primary target dominates the field before PSF subtraction.
}
\end{center}
\end{figure}



\begin{figure}[ht]
 \begin{center}
    \subfigure[]
        {\includegraphics[scale=1.5, angle=0]{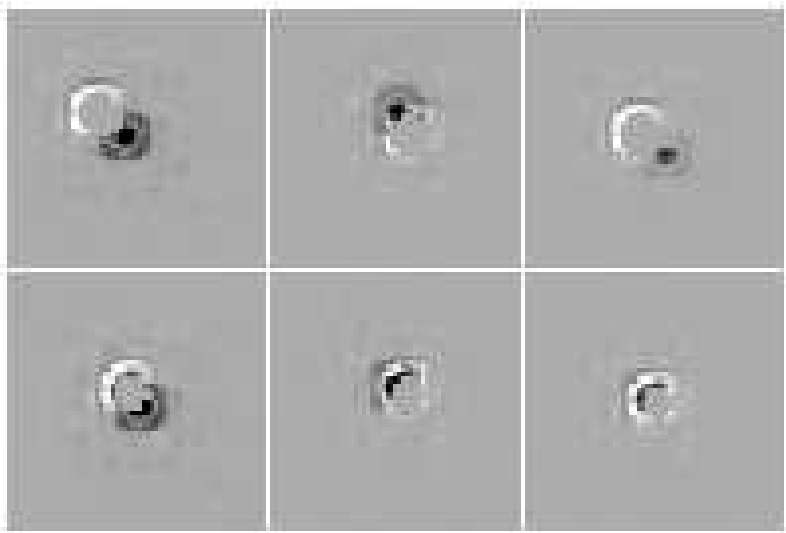}}
    \subfigure[]
	{\includegraphics[scale=0.42, angle=0]{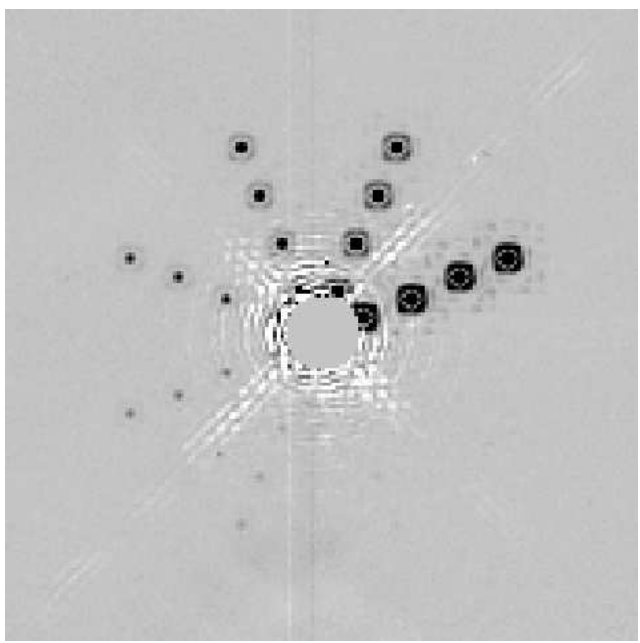}}
    \figcaption[figure5] {Examples of sensitivity simulations around
     GJ 213 (M4.0V, {\it F180M}$=$6.68).  (a) To test sub-arcsecond
     separations, a mosaic of PSF insertions with several separations
     and magnitudes is created.  In this figure the rows represent
     separations of 0\farcs2 (bottom), and 0\farcs4 (top).  The
     columns represent apparent {\it F180M} magnitudes of 9, 10, and
     11 from left to right. The artificial companions are visible at
     all three magnitudes for 0\farcs4 but only at the brightest
     magnitude for 0\farcs2. (b) PSF insertions are laid out in a
     radial pattern to test the sensitivity at separations of 1\farcs0
     and greater. Apparent {\it F180M} magnitudes range from 12 to 19
     in increments of 1, with the rays for 18 and 19 not detectable in
     this case. Separations are 1\farcs0, 2\farcs0, 3\farcs0, and
     4\farcs0. In both simulations the residuals of the PSF
     subtraction are set to zero at a radius interior to the
     artificial companions to facilitate detection. A thorough
     inspection requires using surface and contour plots.  }

\end{center}
\end{figure}


\begin{figure}[ht]
 \begin{center}
    \subfigure[]
	{\includegraphics[scale=0.47, angle=90]{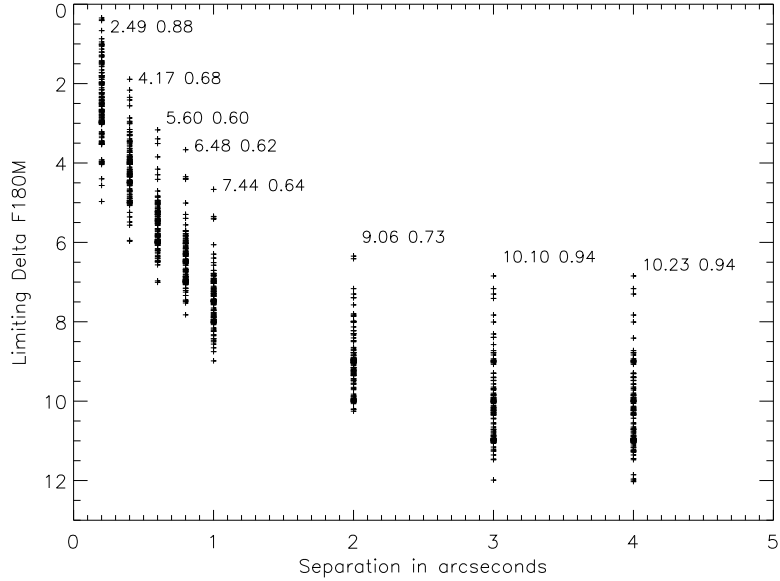}}
    \subfigure[]
	{\includegraphics[scale=0.47, angle=90]{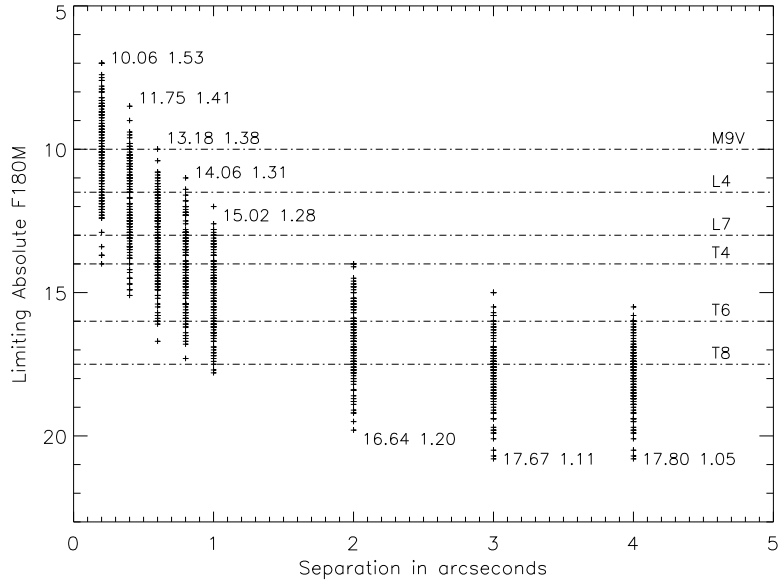}}
    \figcaption[figure6] {Search sensitivities for the eight angular
    separations tested by PSF insertion simulations. In both panels,
    the two numbers next to each cluster of points are the mean and
    standard deviation for that separation, respectively.  (a) The
    ability to detect a companion is primarily determined by the
    angular separation and the components' $\Delta$m.  This
    instrumental representation has a lower standard deviation, but
    does not probe fundamental astrophysical parameters. (b)
    Transforming $\Delta$m into absolute magnitudes yields a range of
    possible companion types detectable at each angular
    separation. The absolute {\it F180M} magnitude for select spectral
    subtypes is taken from the synthetic photometry displayed in
    Figure 3b.  }

\end{center}
\end{figure}



\begin{landscape}
\begin{figure}[ht]
    \subfigure[]
  {\includegraphics[scale=0.45, angle=90]{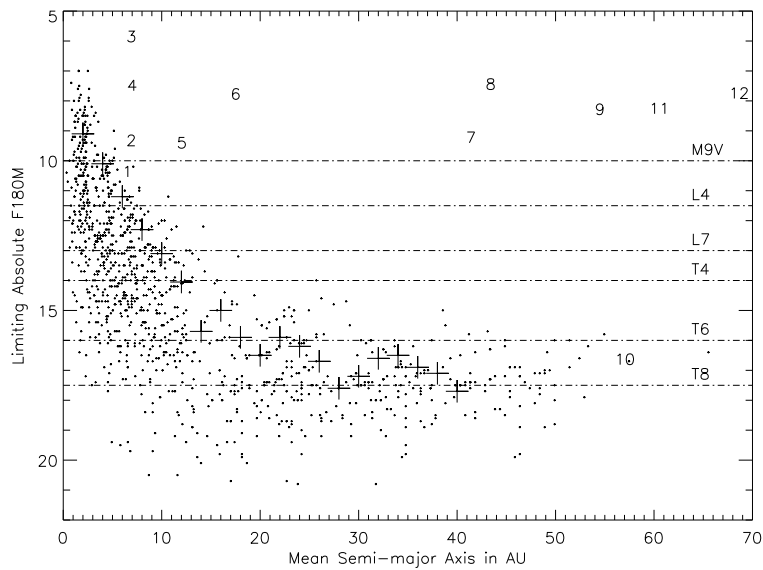}}
    \subfigure[]
  {\includegraphics[scale=0.45, angle=90]{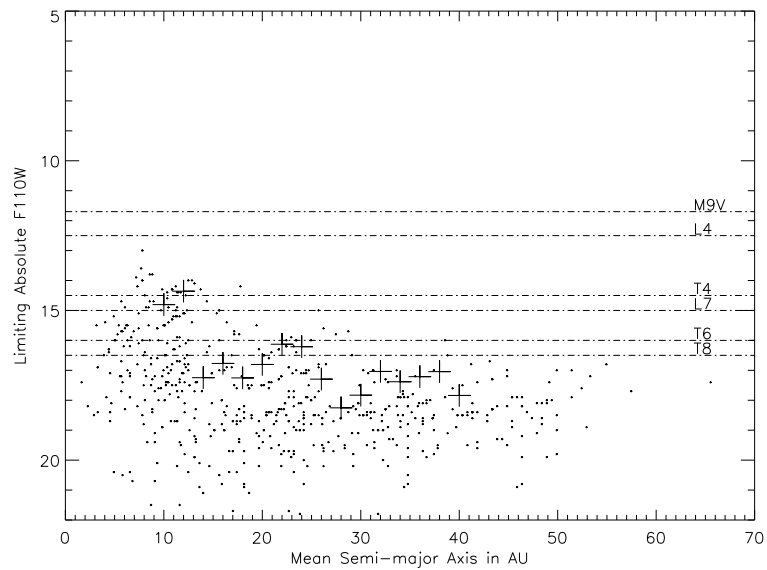}}
  
  \figcaption[figure7] {(a) Search sensitivity displayed as a function
of absolute {\it F180M} magnitude and mean semi-major axis, assuming
$<a>=1.26<\rho>$ \citep{FisherandMarcy1992}. Each dot represents the
sensitivity derived from a PSF insertion around an M dwarf (Table
2). The range of limiting absolute magnitudes is significantly wider
at close separations because all targets were probed at close physical
separations, whereas only targets close to our distance limit of 10 pc
could be probed at wide physical separations given {\it NIC2's} small
field of view. Contrast is also more strongly dependent on overall brightness
at close angular separations. The large plusses represent the absolute magnitude
limits where 90\% of companions can be detected at a given physical
separation. The numbers indicate the positions of the companions listed in Table
5: (1) GJ 84B, (2) GJ 65B, (3) GJ 661B, (4) GJ 257B, (5) GJ 1116B, (6)
GJ 860B, (7) GJ 1245B, (8) GJ 896B, (9) GJ 1230B, (10) GJ 229B, (11)
GJ 618B, and (12) LP 771-95B. The large blank space in the center and
right-hand-side of the diagram is a clear representation of the
``brown dwarf desert''. (b) Same as (a), but using absolute {\it
F110W}, and omitting separations $\leq$ 1\farcs0. While the
sensitivity to T dwarfs is increased in (b), the sensitivity to L
dwarfs is decreased and close separations cannot be probed.
See $\S$5.3 for discussion.}
\end{figure}
\end{landscape}

\begin{figure}[ht]
 \begin{center}
  \includegraphics[scale=0.6, angle=90]{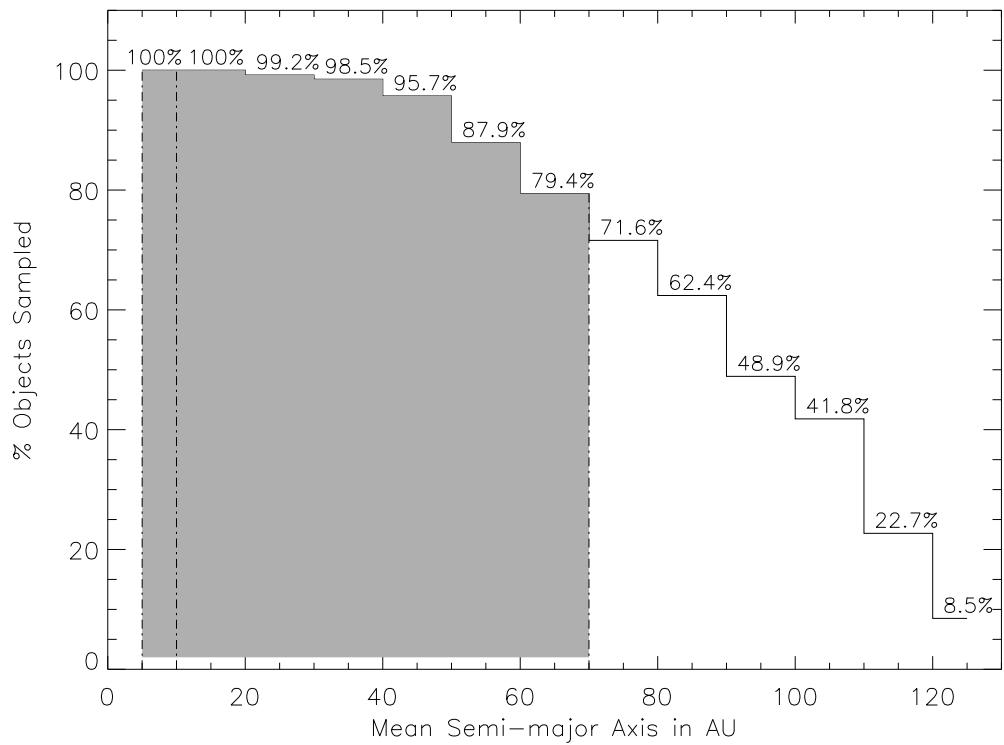}
\figcaption[figure8]{Statistical semi-major axis distribution for
companion search around all 141 M dwarf components within 10 pc.  The
shaded area indicates the separation ranges we consider when
calculating the multiplicity fraction, with the dashed lines
indicating the inner radius limits for M and L dwarfs (5 AU), and for
the T dwarfs (10 AU) and the outer radius for both (70 AU).
Search completeness diminishes with increasing
separation because {\it NIC2's} field of view limits our search radius
to $\sim 9\arcsec$.  }

\end{center}
\end{figure}


\begin{figure}[ht]
 \begin{center}
 \subfigure[]
	{\includegraphics[scale=0.35, angle=90]{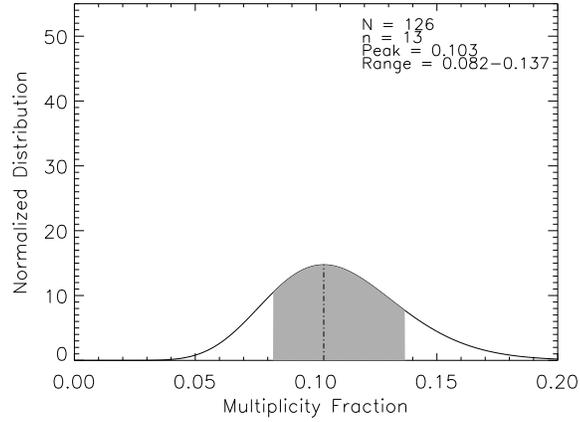}}
    \subfigure[]
	{\includegraphics[scale=0.35, angle=90]{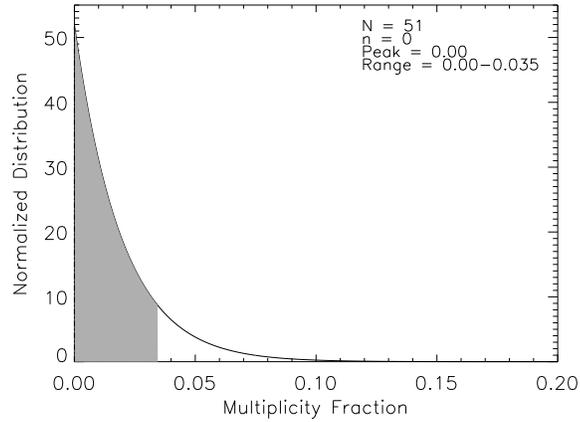}}
    \subfigure[]
	{\includegraphics[scale=0.35, angle=90]{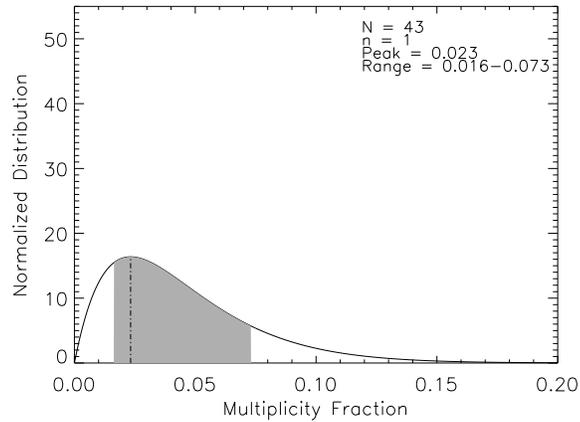}}
\figcaption[figure9]{Probability density distributions for select
multiplicity fractions listed in Table 6, calculated using the
binomial distribution. The shaded areas correspond to 68\% of the area
under the curve, equivalent to the $1\sigma$ confidence range. The
individual plots correspond to: (a) M dwarf companions, (b) L
dwarf companions, (c) L and T dwarf companions.  }

\end{center}
\end{figure}


\begin{figure}[ht]
 \begin{center}
  \includegraphics[scale=0.7, angle=90]{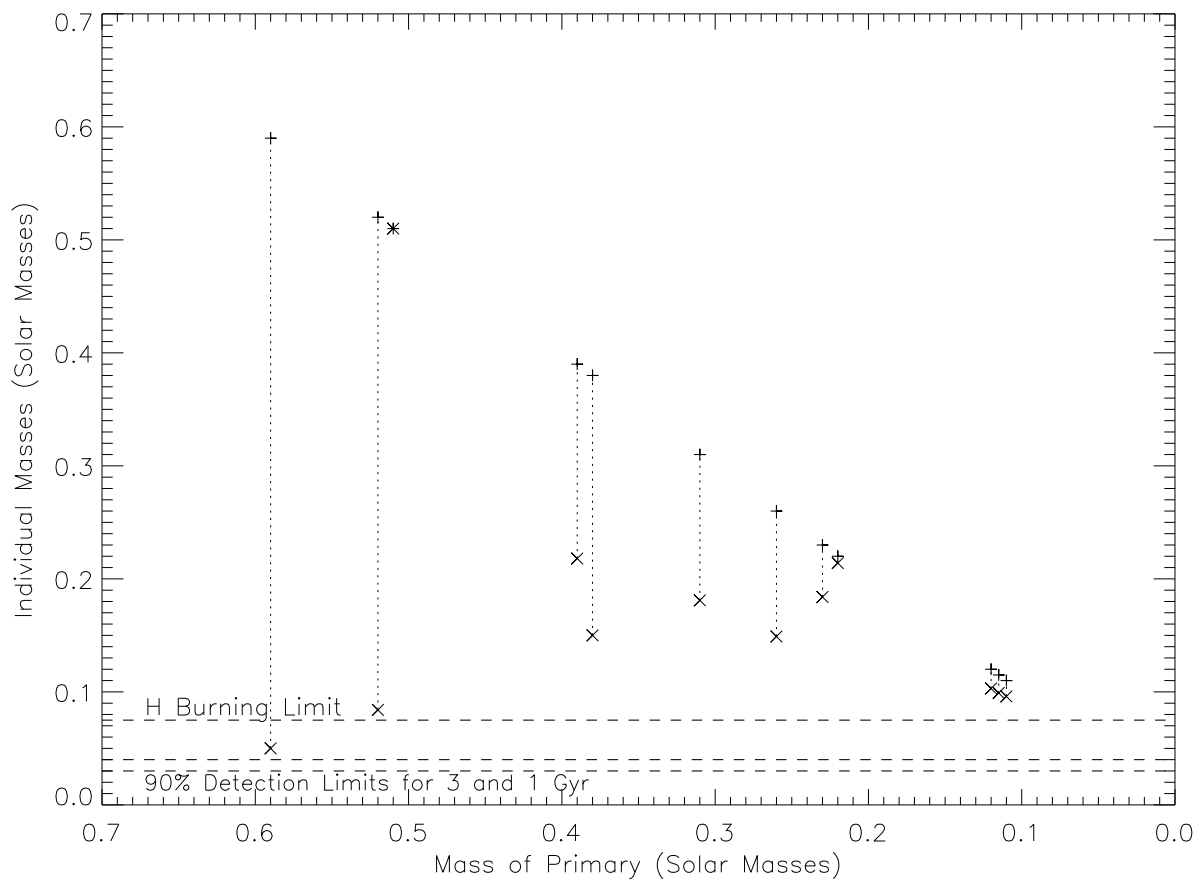}
\figcaption[figure10]{Mass distribution for the binaries in Table 5.
The horizontal dashed lines denote the hydrogen burning limit (0.075
M$_{\sun}$) and the 90\% detection limits for this search assuming
brown dwarf ages of 1 Gyr and 3 Gyr (Table 7). As the masses of the
primary components approach the hydrogen burning limit, the mass
ratios tend to unity, thus implying that brown dwarfs rarely form as
secondaries. See $\S$6.3$-$6.4 for discussion.  From left to right, the
binaries are ordered as they appear in Table 5. }

\end{center}
\end{figure}




\voffset0pt{ \centering
\begin{deluxetable}{lr}
\tabletypesize{\scriptsize}
\tablecaption{Sample Tallies}
\setlength{\tabcolsep}{0.1in}
\label{tab:statistics}
\tablewidth{0pt}

\tablehead{
	   \colhead{Sub-sample }                         &
	   \colhead{Tally}                                }

\startdata
 {\it HST} Visits                                            & 217          \\        
 Total Resolved Targets\tablenotemark{a}                     & 233          \\
 Total Unresolved Targets\tablenotemark{b}                   & 22           \\  
 Total Star Systems\tablenotemark{c}                         & 201          \\
 Resolved Targets within 10 pc\tablenotemark{a}              & 218          \\
 Unresolved Targets within 10 pc\tablenotemark{b}            & 21           \\
 Star Systems within 10 pc\tablenotemark{c}                  & 188          \\
 Resolved M Dwarfs within 10 pc\tablenotemark{a}             & 141          \\  
 Systems with M Dwarf Primary within 10 pc\tablenotemark{c}  & 126          \\  
 Resolved Targets beyond 10 pc\tablenotemark{a}              & 15           \\
 Unresolved Targets beyond 10 pc\tablenotemark{b}            & 1            \\
 Star Systems beyond 10 pc\tablenotemark{c}                  & 13           \\
 
\enddata
\tablenotetext{a}{Each PSF not known to be a blend of more than one 
                  star or brown dwarf and that is not a background source is counted as one ``resolved target''.}
\tablenotetext{b}{Each PSF known to be the blend of two or more stars or brown dwarfs in the same system
                  is counted as one ``unresolved target''.}
\tablenotetext{c}{Denotes known physical association at any physical separation, including systems comprising
                  multiple fields of view.}

\end{deluxetable}



\begin{landscape}
\voffset0pt{\centering
\begin{deluxetable}{lclrlrrlrrrrrrl}
\tabletypesize{\scriptsize}
\tablecaption{Sensitivity to Companions}
\setlength{\tabcolsep}{0.1in}
\tablewidth{0pt}
\tablehead{
           \colhead{Name    }            &
           \colhead{Resolved PSFs}     &
           \colhead{Spectral }    & 
           \colhead{Distance}         & 
           \colhead{Epoch}            &
           \colhead{App. Mag.}      &
	   \multicolumn{8}{c}{Absolute {\it F180M} Magnitude Limit} \\

	   \colhead{}              &           
           \colhead{in FOV\tablenotemark{1}}        &
           \colhead{Type}              &
           \colhead{(pc)}          &
           \colhead{}              &
	   \colhead{({\it F180M})} &
	   \colhead{0\farcs2}     &
           \colhead{0\farcs4}     &
           \colhead{0\farcs6}     &
           \colhead{0\farcs8}     &
           \colhead{1\farcs0}     &
           \colhead{2\farcs0}     &
           \colhead{3\farcs0}     &
           \colhead{4\farcs0}     &	   
	   \colhead{Notes}              }

\startdata
          GJ 915 & \nodata  &     DA5 &   8.1 &   2003 Jun  3 &  12.53 &   11.9 &   12.4 &   14.4 &   14.4 &   14.9 &   15.4 &   15.4 &   15.4 &    \tablenotemark{     a} \\
        GJ 1001A & A        &   M3.0V &  13.0 &   1998 Aug  3 &   7.97 &    8.4 &   11.9 &   13.4 &   13.9 &   14.4 &   16.4 &   16.4 &   16.4 &    \tablenotemark{     b,g} \\
        GJ 1     & \nodata  &   M1.5V &   4.3 &   1998 Jan 10 &   4.64 &    8.0 &   11.0 &   12.5 &   13.0 &   14.5 &   16.5 &   17.5 &   18.5 &    \nodata \\
        GJ 1002  & \nodata  &   M5.0V &   4.6 &   2002 Oct 27 &   7.79 &   12.4 &   13.4 &   14.9 &   15.9 &   16.4 &   18.4 &   19.4 &   19.4 &    \nodata \\
      GJ 1005AB  & AB       &  M3.5VJ &   5.9 &   2002 Oct  3 &   6.71 &   10.1 &   12.1 &   13.1 &   14.1 &   15.1 &   17.1 &   18.1 &   18.1 &    \tablenotemark{   b,d,f} \\
        GJ 15A   & A        &   M1.5V &   3.5 &   2003 Jun 26 &   4.48 &  \nodata   &   10.7 &   12.2 &   12.2 &   13.2 &   15.2 &   16.2 &   16.2 &    \tablenotemark{    b,d,i} \\
        GJ 15B   & B        &   M3.5V &   3.5 &   1998 Aug 19 &   6.20 &   12.4 &   12.4 &   14.4 &   15.4 &   15.4 &   17.4 &   18.4 &   18.4 &    \tablenotemark{    b} \\
         GJ 17   & \nodata  &   F9.5V &   8.5 &   1997 Aug 29 &   2.81 &  \nodata   &    8.1 &    9.1 &   10.1 &   10.6 &   13.1 &   14.1 &   15.1 &    \tablenotemark{     i} \\
\enddata
\tablenotetext{1}{A/B denotes that components A and B are resolved. AB denotes an unresolved known multiple.}
\tablenotetext{a}{No PSF subtraction}
\tablenotetext{b}{Known multiple system, excluding planets}
\tablenotetext{c}{Sensitivity measured around unresolved or very close binary}
\tablenotetext{d}{{\it 2MASS H} magnitude for primary}
\tablenotetext{e}{Off-center by more than 5$\arcsec$}
\tablenotetext{f}{Sensitivity measured around brighter component only}
\tablenotetext{g}{Beyond 10 pc}
\tablenotetext{h}{Sensitivity measurements exclude $30\degr$  between components.}
\tablenotetext{i}{Saturated core}
\end{deluxetable}
}
\end{landscape}



\voffset0pt{ \centering
\begin{deluxetable}{lrrll}
\tabletypesize{\scriptsize}
\tablecaption{Background Sources with Companion-Like Colors}
\setlength{\tabcolsep}{0.09in}
\label{tab:statistics}
\tablewidth{0pt}

\tablehead{
	   \colhead{Primary}                      &
	   \colhead{Separation}                   &
           \colhead{P. A. (deg)}                  &
 	   \colhead{Mimics}                       & 
	   \colhead{Deciding Factor                }}
\startdata

 GJ 633  &  8\farcs5 & 157.3  & Late L in {\it F207M-F222M}  &  Too red to be late L in {\it F110W-F180M}         \\ 
 GJ 633  &  3\farcs3 & 268.6  & Late L in {\it F207M-F222M}  &  Too red to be late L in {\it F110W-F180M}         \\
 GJ 1093 &  8\farcs0 & 250.6  & Mid M in {\it F110W} colors  &  Background M1 or earlier in {\it F207M-F222M}     \\
 GJ 1224 & 12\farcs6 &  91.4  & Mid L in {\it F110W-F222M}   &  Background early M in {\it F180M-F222M}           \\
 GJ 367  &  8\farcs4 &   3.3  & Early L in {\it F180M-F207M} &  Background main sequence in {\it F110W-F222M}     \\  
 GJ 438  & 13\farcs2 & 254.0  & Hot white dwarf              &  Too red to be hot WD in {\it F110W-F180M}         \\

\enddata
\end{deluxetable}
}


\voffset0pt{ \centering
\begin{deluxetable}{lccccrrrrr}
\tabletypesize{\scriptsize}
\tablecaption{Astrometry of Unsaturated Resolved Systems}
\setlength{\tabcolsep}{0.05in}
\label{tab:statistics}
\tablewidth{0pt}

\tablehead{
            \colhead{Pair}                            &
	    \colhead{$\rho$}                          &
	    \colhead{$\sigma_{\rho}$}                 &
            \colhead{P.A.}                            &
            \colhead{$\sigma_{P.A.}$}                 &
	    \colhead{Epoch}                           &
            \multicolumn{4}{c}{$\Delta$ Magnitude}    \\
            \colhead{}                                &
            \colhead{($\arcsec$)}                     &
            \colhead{}                                &
            \colhead{($\degr$ E of N)}                &
            \colhead{}                                &
            \colhead{}                                &
            \colhead{{\it F110W}\tablenotemark{a}}    &
            \colhead{{\it F180M}}                     &
            \colhead{{\it F207M}}                     &
            \colhead{{\it F222M}}                     }
\startdata

GJ 1005AB    &   0.329  & 0.008  & 234.4  & 1.0  &      2002.7532 &  \nodata &  1.27 & 1.31 &  1.32  \\
GJ 65AB      &   1.653  & 0.008  & 103.3  & 0.2  &      2002.8540 &  \nodata &  0.18 & 0.15 &  0.16  \\
GJ 84AB\tablenotemark{b}   &   0.443  & 0.006  & 103.4  & 1.0  &      2002.7506 &  4.59    &  4.01 & 4.18 &  3.82  \\
GJ 105AC     &   3.220  & 0.036  & 293.5  & 0.4  &      1998.0225 &  5.96    &  5.87 & 5.45 &  5.40  \\
LP 771-95AB  &   7.706  & 0.008  & 315.0  & 0.0  &      2003.4620 &  \nodata &  0.36 & 0.36 &  0.35  \\
LP 771-95BC  &   1.344  & 0.008  & 138.1  & 0.2  &      2003.4620 &  \nodata &  0.56 & 0.54 &  0.50  \\
GJ 169.1AB   &   9.201  & 0.008  &  63.7  & 0.0  &      2002.7693 &  \nodata &  5.89 & 5.98 &  6.07  \\
GJ 229AB     &   7.627  & 0.031  & 164.1  & 0.1  &      1997.6202 &  9.63    & 11.44 & 9.44 & 10.75  \\
\enddata
\tablenotetext{a}{Missing values correspond to stars with central core saturation for which a PSF fit
is not available.}
\tablenotetext{b}{Values from G04, Table 4.} 
\end{deluxetable}
}
 


\voffset0pt{ \centering
\begin{deluxetable}{llccllcccc  }
\tabletypesize{\scriptsize}
\tablecaption{M Dwarf Multiple Systems with Separations Between 5 and 70 AU Recovered in the Survey}
\setlength{\tabcolsep}{0.03in}
\label{tab:statistics}
\tablewidth{0pt}

\tablehead{
	   \colhead{A Component\tablenotemark{a}}     &
           \colhead{Spectral }                        &
           \colhead{$M_{H}$\tablenotemark{b}}         &
           \colhead{Mass\tablenotemark{c}}            &                   
 	   \colhead{B Component}                      & 
           \colhead{Spectral }                        &
           \colhead{$M_{H}$\tablenotemark{b}}         &
           \colhead{Mass\tablenotemark{c}}            &
           \colhead{Angular}                          &
           \colhead{Inferred }         \\
           \colhead{ }                                &
           \colhead{Type}                             &
           \colhead{ }                                &
           \colhead{M$_{\sun}$ }                      &
           \colhead{ }                                &
           \colhead{Type}                             &
           \colhead{ }                                &
           \colhead{M$_{\sun}$ }                      &
           \colhead{Separation }                      &
           \colhead{Semi-major Axis (AU)\tablenotemark{f} }            }
            
\startdata
GJ  229A   & M1.5V  & 5.41   & 0.59   &  GJ  229B   & T6.0   & 16.81   & 0.05:\tablenotemark{d}   &  7\farcs62  & 55.2   \\
GJ   84A   & M2.5V  & 5.95   & 0.52   &  GJ   84B   & M7.0V  & 10.54   & 0.084\tablenotemark{e}   &  0\farcs44  &  5.2   \\ 
GJ  661A   & M3.0V  & 6.03   & 0.51   &  GJ  661B   & M3.0V  &  6.03   & 0.51   &  0\farcs70  &  5.6   \\
GJ  896A   & M3.5V  & 6.60   & 0.39   &  GJ  896B   & M4.5V  &  7.62   & 0.22   &  5\farcs35  & 42.1   \\ 
GJ  618A   & M2.5V  & 6.62   & 0.38   &  GJ  618B   & M4.5V  &  8.43   & 0.15   &  5\farcs62  & 58.7   \\  
GJ  860A   & M3.0V  & 7.02   & 0.31   &  GJ  860B   & M4.0V  &  7.95   & 0.18   &  3\farcs19  & 16.2   \\
GJ 1230A   & M4.0V  & 7.34   & 0.26   &  GJ 1230B   & M5.0V  &  8.46   & 0.15   &  5\farcs11  & 53.2   \\
LP 771-95A & M2.5V  & 7.56   & 0.23   &  LP 771-95B & M3.5V  &  7.92   & 0.18   &  7\farcs74  & 67.8   \\
\enddata
\tablenotetext{a}{Ordered by decreasing mass, as shown in Figure 9.}
\tablenotetext{b}{{\it H} band photometry from {\it 2MASS}. Close binaries were deconvolved adopting $\Delta H = \Delta F180M$.}
\tablenotetext{c}{Based on the Mass-Luminosity Relation of \citet{HenryandMcCarthy1993}.}
\tablenotetext{d}{Estimate based on \citet{Allardetal1996}.}
\tablenotetext{e}{G04}
\tablenotetext{f}{statistically corrected for projection effects \citep{FisherandMarcy1992}.}

\end{deluxetable}



\voffset0pt{ \centering
\begin{deluxetable}{ccccccc}
\tabletypesize{\scriptsize}
\tablecaption{Summary of M Dwarf Multiplicity Fractions}
\setlength{\tabcolsep}{0.1in}
\label{tab:statistics}
\tablewidth{0pt}
\tablehead{
           \colhead{Companion Range}                   &
           \colhead{Search Radius  }                   &
           \colhead{Systems        }                   &  
           \colhead{Detections     }                   &
           \colhead{Hidden         }                   &
           \colhead{Mult. Fraction }                   & 
           \colhead{Volume         }                   \\
           \colhead{Spectral Type}                     &
           \colhead{AU}                                &
           \colhead{Probed}                            &      
           \colhead{}                                  &
           \colhead{}                                  &
           \colhead{\%}                                &
           \colhead{Limited?}                          }
\startdata
M0V - M9V  &  5 - 70  & 126  &  11  &  2       & $10.3_{-2.1}^{+3.4}$   & Yes   \\ 
L0 - L9    &  5 - 70  &  51  &  0   &  0       & $0.0_{-0.0}^{+3.5}$    & No    \\ 
L0 - L9    & 12 - 70  & 126  &  0   &  0       & $0.0_{-0.0}^{+1.4}$    & Yes   \\
L0 - T9    & 10 - 70  &  43  &  1   &  0       & $2.3_{-0.7}^{+5.0}$    & No    \\ 
L0 - T9    & 14 - 70  & 126  &  1   &  0       & $0.8_{-0.2}^{+1.8}$    & Yes   \\ 
\enddata
\end{deluxetable}
}



\voffset0pt{ \centering
\begin{deluxetable}{cccccc}
\tabletypesize{\scriptsize}
\tablecaption{Brown Dwarf Masses (M$_{\sun}$) Based on Models of \citet{Chabrieretal2000}.}
\setlength{\tabcolsep}{0.1in}
\label{tab:statistics}
\tablewidth{0pt}

\tablehead{
	   \colhead{ }                                &
	   \colhead{L3}                               &
 	   \colhead{L5}                               & 
	   \colhead{L8}                               &
           \colhead{T5}                               & 
           \colhead{T7}                              \\
           \colhead{T$_{eff}$\tablenotemark{a}}       &
           \colhead{2000}                             &
           \colhead{1750}                             &
           \colhead{1500}                             &
           \colhead{1200}                             &
           \colhead{ 900:}                            }

\startdata
1 Gyr  &   0.070  &  0.060  &  0.050  &  0.050 &  0.030: \\
3 Gyr  &   0.073  &  0.070  &  0.057  &  0.052 &  0.040: \\
5 Gyr  &   0.075  &  0.072  &  0.065  &  0.065 &  0.050: \\
\enddata
\tablenotetext{a}{\citet{Golimowskietal2004a}}
\end{deluxetable}

\end{document}